\batchmode
\makeatletter
\def\input@path{{"C:/Users/marfon/OneDrive - The University of Liverpool/Papers/PMBM_on_tracklets/"}}
\makeatother
\documentclass[onecolumn,british,twocolumn]{IEEEtran}
\usepackage[T1]{fontenc}
\usepackage[latin9]{inputenc}
\usepackage{array}
\usepackage{float}
\usepackage{multirow}
\usepackage{tikz}
\usepackage{amsmath}
\usepackage{amsthm}
\usepackage{amssymb}
\usepackage{graphicx}
\usepackage{rotating}
\PassOptionsToPackage{normalem}{ulem}
\usepackage{ulem}

\makeatletter

\providecommand{\tabularnewline}{\\}
\floatstyle{ruled}
\newfloat{algorithm}{tbp}{loa}
\providecommand{\algorithmname}{Algorithm}
\floatname{algorithm}{\protect\algorithmname}
\providecolor{lyxadded}{rgb}{0,0,1}
\providecolor{lyxdeleted}{rgb}{1,0,0}
\usetikzlibrary{calc}
\newcommand{\lyxobjectsout}[1]{%
  \bgroup%
  \color{lyxdeleted}%
  \tikz{
    \node[inner sep=0pt,outer sep=0pt](lyxdelobj){#1};
    \draw($(lyxdelobj.south west)+(2em,.5em)$)--($(lyxdelobj.north east)-(2em,.5em)$);
  }
  \egroup%
}
\DeclareRobustCommand{\lyxdisplayobjdeleted}[4][]{%
  \ifx#4\empty\else%
     \leavevmode\\%
     \lyxobjectsout{\parbox{\linewidth}{#4}}%
  \fi%
}
\DeclareRobustCommand{\lyxudisplayobjdeleted}[4][]{%
  \ifx#4\empty\else%
     \leavevmode\\%
     \raisebox{-\belowdisplayshortskip}{%
                \lyxobjectsout{\parbox[b]{\linewidth}{#4}}}%
     \leavevmode\\%
  \fi%
}

\theoremstyle{plain}
\newtheorem{thm}{\protect\theoremname}
\theoremstyle{plain}
\newtheorem{lem}[thm]{\protect\lemmaname}

\usepackage{cite} 
\usepackage[margin=8pt,font=footnotesize]{caption}
\usepackage{algorithm}
\newcommand{\StatexIndent}[1][3]{%
  \setlength\@tempdima{\algorithmicindent}%
  \Statex\hskip\dimexpr#1\@tempdima\relax}
\usepackage{algpseudocode}
\algdef{S}[FORALL]{ForAllNoDo}[1]{\algorithmicforall #1}%
\algdef{S}[FOR]{ForNoDo}[1]{\algorithmicfor\ #1}%

\DeclareMathOperator{\diag}{diag}
\usepackage{cases}
\usepackage{empheq}
\usepackage{amsmath}  % You need this for the math
\allowdisplaybreaks

\ifdefined\showcaptionsetup
 \PassOptionsToPackage{caption=false}{subfig}
\fi
\usepackage{subfig}
\makeatother

\usepackage{babel}
\providecommand{\lemmaname}{Lemma}
\providecommand{\theoremname}{Theorem}

\begin{document}
\title{Poisson multi-Bernoulli mixture filter for trajectory measurements}
\author{Marco Fontana, Ángel F. García-Fernández, Simon Maskell\thanks{M. Fontana and S. Maskell are with the Department of Electrical Engineering and Electronics, University of Liverpool, Liverpool L69 3GJ, United Kingdom (emails: \{marco.fontana, s.maskell\}@liverpool.ac.uk).\\
Á. F. García-Fernández was with the Department of Electrical Engineering and Electronics, University of Liverpool, Liverpool L69 3GJ, United Kingdom. He is now with the IPTC, ETSI de Telecomunicaci\'on, Universidad Politécnica de Madrid, Av. Complutense 30, 28040, Madrid, Spain (e-mail: \mbox{angel.garcia.fernandez@upm.es}).}}
\maketitle
\begin{abstract}
This paper presents a Poisson multi-Bernoulli mixture (PMBM) filter
for multi-target filtering based on sensor measurements that are sets
of trajectories in the last two-time step window. The proposed filter,
the trajectory measurement PMBM (TM-PMBM) filter, propagates a PMBM
density on the set of target states. In prediction, the filter obtains
the PMBM density on the set of trajectories over the last two time
steps. This density is then updated with the set of trajectory measurements.
After the update step, the PMBM posterior on the set of two-step trajectories
is marginalised to obtain a PMBM density on the set of target states.
The filter provides a closed-form solution for multi-target filtering
based on sets of trajectory measurements, estimating the set of target
states at the end of each time window. Additionally, the paper proposes
computationally lighter alternatives to the TM-PMBM filter by deriving
a Poisson multi-Bernoulli (PMB) density through Kullback-Leibler divergence
minimisation in an augmented space with auxiliary variables. The performance
of the proposed filters are evaluated in a simulation study.
\end{abstract}


\section{Introduction\label{sec:tracklets:Introduction}}

Multi-target filtering is the algorithmic process of continuously
estimating a time-evolving set of target states in a stochastic environment,
based on incoming noisy data streams \cite{Blackman91,Maskell2009,BarShalom2011}.
It has widespread applications, including radar and sonar tracking,
autonomous vehicle navigation, video surveillance, and biological
studies \cite{Skolnik1962,Choi2013,Maggio2008,Chenouard2013}.

Some of the most widely used approaches include multiple hypothesis
tracking (MHT) \cite{Reid79}, joint probabilistic data association
(JPDA) \cite{Fortmann83}, and random finite sets (RFS) \cite{Mahler14}.
Finite Set Statistics (FISST) can be used to provide a general Bayesian
framework for multi-target filtering problems involving an unknown
and time-varying number of targets \cite{Goutsias2012,Goodman2013}.
This approach models both multi-target states and multi-target measurements
as RFSs under a rigorous unified framework \cite{Mahler14}, which
is closely related to point process theory \cite{Schlangen2018}.
An early approximate solution to the multi-target filtering problem
based on FISST is the Probability Hypothesis Density (PHD) filter
\cite{Mahler2003}. More recently, RFS-based filters that generate
explicit data association hypotheses have been introduced, such as
the Poisson multi-Bernoulli mixture (PMBM) filter \cite{Garcia-Fernandez18},
the multi-Bernoulli mixture (MBM) filter \cite{Garcia-Fernandez18,GarciaFernandez2019}
and the generalised labelled multi-Bernoulli (GLMB) \cite{Vo13}.
Finally, in recent years, methods aimed at estimating sets of target
trajectories directly from the posterior density on the set of trajectories
have been proposed \cite{GarciaFernandez2019a,GarciaFernandez2020}.

Standard multi-target filtering algorithms are typically designed
for detectors that provide information on the set of targets at specific,
discrete points in time. The standard point-target measurement model
\cite[Sec. 12.3]{Mahler07}, as well as the extended target \cite{Granstroem2017}
and track-before-detect \cite{Davey2008} measurement models commonly
found in the literature are based on this assumption. Nevertheless,
certain sensors may generate measurements that span a time interval
during which the target may have moved.

For instance, the detector for distributed acoustic sensing proposed
in \cite{Fontana2025} operates within a time window producing detections
that are trajectories. Each of these trajectory measurements may have
been generated by a target or by clutter. Trajectory measurements
can also be generated by some track-before-detect (TkBD) processing
methods, which delay target declaration until enough data has been
accumulated \cite{Moyer2011}. Various batch processing algorithms,
such as envelope interpolation, maximum likelihood estimation \cite{Tonissen1998},
and dynamic programming \cite{Barniv1987}, also follow this approach.
When integration periods are extended with minimal acceleration, individual
scans form straight lines in Cartesian-time or range-Doppler frequency-time
space \cite{Carlson1994}, detectable using techniques like the Hough
transform \cite{Hough,Duda1972}, RANSAC \cite{Fischler1981}, and
Expectation-Maximization \cite{Dempster1977}. Each line can be characterised
by estimating the initial position and velocity of the target \cite{Arnold1984}.

In space surveillance, fast-moving satellites may appear as streaks
(trajectory measurements) in telescope images, particularly in low
SNR conditions \cite{Levesque2007,Finelli2017}. A method for tracking
streaking targets in video frames using the Maximum Likelihood Probabilistic
Multi-Hypothesis Tracker (MLPMHT) \cite{Schoenecker2011} and a target
state defined by the starting and ending positions in each batch was
proposed in \cite{Finelli2017}.

Trajectory measurements may also arise in a multi-sensor scenario,
where multiple measurements, presumably for the same target, are combined
at the sensor level of a `composite measurement fusion' architecture
\cite{Drummond1997}. These trajectory measurements, or the tracklets
derived from them \cite[Ch. 10]{Blackman91}, are subsequently used
to update the global tracks, which are maintained at the central level
of the system.

\begin{figure}
\centering{}\includegraphics[scale=0.35]{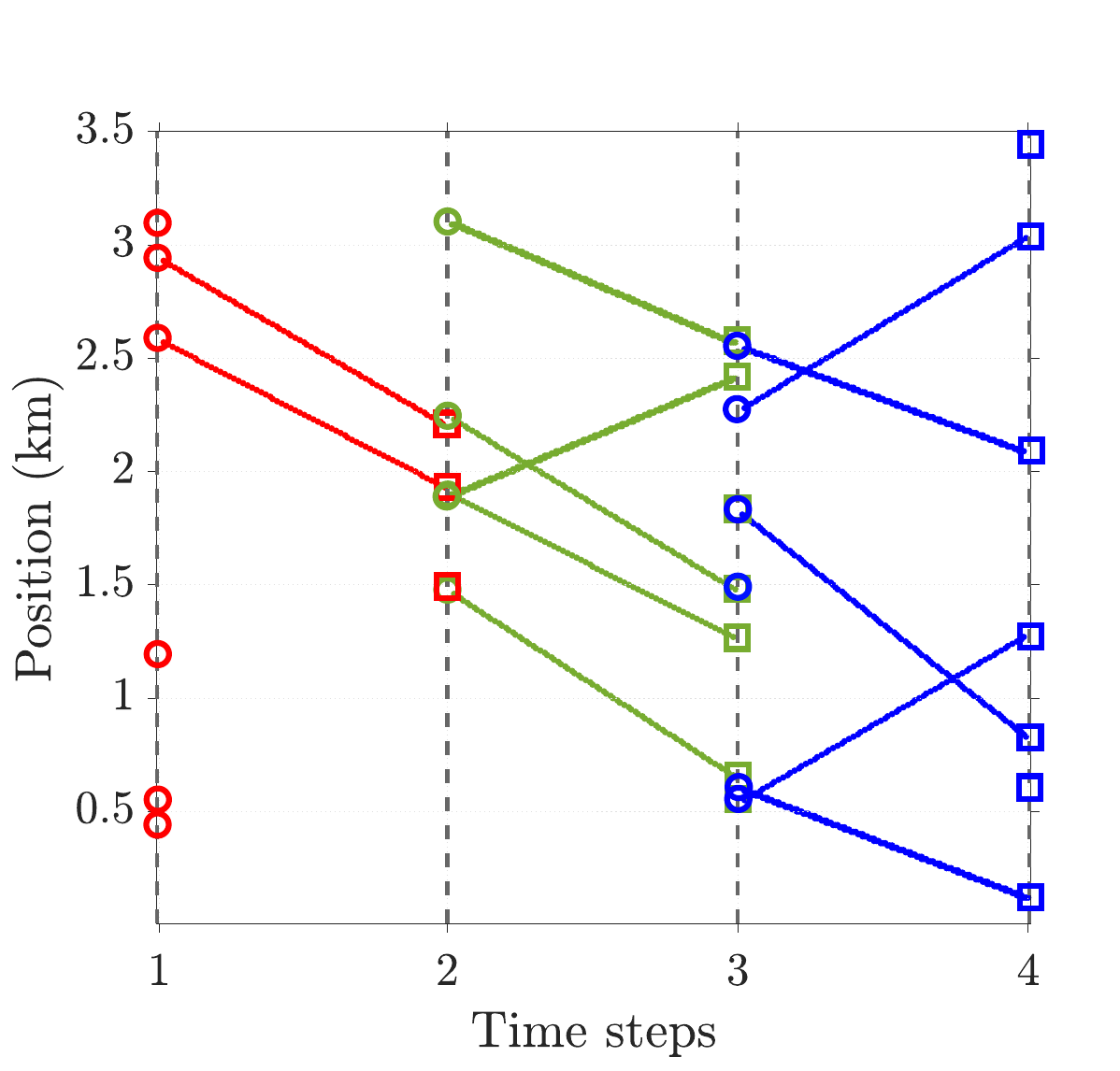}\caption{An example of three sets of trajectory measurements across three time
windows, spanning time steps 1 to 4. Each set of trajectory measurements
is represented by a distinct colour. The vertical dashed lines indicate
the start and end of each time window. Each trajectory measurement
is either a trajectory connecting two one-dimensional point detections
at time step $k$ (circle) and $k+1$ (square), a point detection
at time step $k$ or a point detection at time step $k+1$. \label{fig:tracklets:example_traj_measurements}}
\end{figure}

Motivated by the need for multi-target filters capable of processing
trajectory measurements, we develop a multi-target PMBM filter, referred
to as trajectory measurement PMBM (TM-PMBM), that provides the closed-form
recursion for computing the posterior distribution of the set of targets
given a sequence of sets of trajectory measurements, each lasting
up to two time steps. These measurement trajectories can represent
point detections at time steps $k$ and $k+1$, and also straight
lines between time steps $k$ and $k+1$, as illustrated in Fig. \ref{fig:tracklets:example_traj_measurements}.
In this paper, each measurement of this type is a trajectory that
can last up to two time steps, and is referred to as trajectory measurement.
The set of received measurements at each time step is therefore a
set of trajectories \cite{GarciaFernandez2020}, comprising both target-generated
trajectory measurements and clutter trajectory measurements. Note
that, whereas existing trajectory PMBM filters \cite{GarciaFernandez2019a,GarciaFernandez2020,Granstroem2019}
estimate trajectories from point measurements, the proposed filter
derives a closed-form PMBM recursion that directly processes raw two-time
step trajectory observations obtained at the sensor level.

\begin{figure}
\centering{}\includegraphics[scale=0.35]{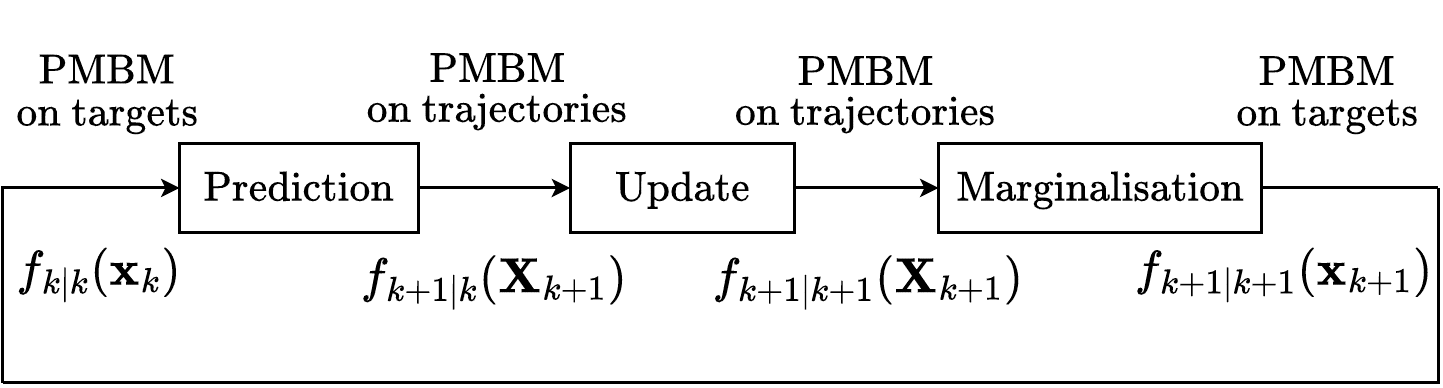}\caption{Diagram of the PMBM filter based on trajectory measurements, the TM-PMBM
filter. After the prediction from time step $k$, the TM-PMBM filter
incorporates information on the set of trajectories at time steps
$k$ and $k+1$, resulting in a PMBM density on this set of trajectories.
The trajectory measurements at time step $k+1$, which can span the
time steps $k$ and $k+1$, are then used to update the PMBM density
on the set of trajectories at time steps $k$ and $k+1$. A marginalisation
step is then done to keep the information on the set of targets at
time step $k+1$. This step also results in PMBM density.\label{fig:tracklets:filter-schematic}}
\end{figure}

Fig. \ref{fig:tracklets:filter-schematic} shows a schematic of the
proposed filter. At each time window, spanning from time step $k$
to $k+1$, the prediction step is applied to the posterior PMBM density
on the set of target states at time step $k$, resulting in a PMBM
density according to the trajectory PMBM filter prediction step \cite{Granstroem2019}.
The resulting predicted PMBM density is defined over two-step trajectories
which, like the trajectory measurements, consist of up to two target
states: one at time step $k$ and one at $k+1.$ Since each trajectory
measurement in the current time window can contain information on
a target trajectory from time step $k$ to $k+1$, the update step
in this interval must be applied on the density that contains information
on the target trajectories from time step $k$ to $k+1$, resulting
in the posterior PMBM density over the two-step trajectories. Finally,
the marginalisation step retains only the information on the set of
targets at time step $k+1$, resulting in the PMBM posterior density
on the target states. This paper also proposes the Gaussian implementation
of the TM-PMBM filter, which is trajectory-measurement version of
the (track-oriented) PMB filter in \cite{Williams15a} and can be
derived via KLD minimisation with auxiliary variables \cite{GarciaFernandez2020a}.

The remainder of the paper is organised as follows. In Section \ref{sec:tracklets:prob_formulation},
we outline the multitarget filtering problem by defining the variables
and models for trajectory measurements and two-step (target) trajectories.
In Section \ref{sec:tracklets:filtering_recursion}, we derive the
TM-PMBM filtering recursion, outlining the prediction, update and
marginalisation steps. The Gaussian implementation of the proposed
TM-PMBM filter is presented in Section \ref{sec:tracklets:Gaussian-TM-PMBM-filter},
while Section \ref{sec:tracklets:TM-PMB-filter} describes the TM-PMB
filter. Finally, Section \ref{sec:tracklets:Simulations} presents
the simulation results, followed by the conclusions drawn in Section
\ref{sec:tracklets:Conclusions}.

\section{Problem formulation\label{sec:tracklets:prob_formulation}}

We develop the multi-target PMBM filter where, at each time step,
the set of measurements contains trajectories in the last two time
steps. Similarly to the point-target measurement model \cite{Mahler14},
each target can generate at most one target-generated trajectory measurement.
The set of trajectory measurements also contains clutter. To introduce
this problem, we first explain the required variables and notation
involved.

In Section \ref{subsec:tracklets:Variables}, we define the notation
used to represent target states, two-step (target) trajectories and
trajectory measurements, along with their corresponding spaces. Section
\ref{subsec:tracklets:trans_model_gen} reviews the standard transition
model and the PPP birth model, extending them to the context of the
two-step trajectories. Section \ref{subsec:tracklets:meas_model}
introduces the measurement model we propose for two-step trajectories.
Finally, Section \ref{subsec:PMBM-solution-overview} revisits the
PMBM posterior density over the set of targets and extends it to the
set of two-step trajectories.

\subsection{Variables\label{subsec:tracklets:Variables}}

\subsubsection{Single-target state and set of targets\label{subsec:Single-target-state-and}}

In the context of multi-target systems, we regard filtering as the
estimation of the states of a time-varying number of targets at the
current time step $k\in\mathbb{N}^{+}$. We denote a single-target
state as $x_{k}\in\mathbb{R}^{n_{x}}$, and the set of target states
at time $k$ as $\mathbf{x}_{k}\in\mathcal{F}(\mathbb{R}^{n_{x}})$,
where $\mathcal{F}(\mathbb{R}^{n_{x}})$ is the set of all finite
subsets of $\mathbb{R}^{n_{x}}$. The set $\mathbf{x}_{k}$ is modelled
as a random finite set, meaning that both the cardinality $|\mathbf{x}_{k}|$
and the elements of the set (i.e., the target states) are random variables
\cite{Mahler14}.

\subsubsection{Single-trajectory state and set of trajectories over $k$ and $k+1$\label{subsec:Single-trajectory-state-and}}

We consider that the time window $k+1$ is defined by two consecutive
time steps $k$ and $k+1$. The space of the single target trajectories
in this time window, is denoted by $T_{(k+1)}$. That is, $T_{(k+1)}$
is the disjoint union of three distinct spaces. These spaces represent
the following cases: there is a single target present at time step
$k$ but not at time step $k+1$, there is a single target born at
time step $k+1$, and there is a single target present at both time
steps $k$ and $k+1$. That is, mathematically, we can write the single-trajectory
space for time steps $k$ and $k+1$ as \cite{GarciaFernandez2020}
\begin{align}
T_{(k+1)} & =T_{(k+1)}^{1}\uplus T_{(k+1)}^{2}\uplus T_{(k+1)}^{3}\nonumber \\
 & =\left[\{k\}\times\mathbb{R}^{n_{x}}\right]\uplus\left[\{k+1\}\times\mathbb{R}^{n_{x}}\right]\uplus\left[\{k\}\times\mathbb{R}^{2n_{x}}\right],\label{eq:tracklets:traj_space}
\end{align}
where $\uplus$ stands for disjoint union. Given a trajectory $\left(t,x_{1:\nu}\right)\in T_{(k+1)}$,
$t$ represents the initial time step of the single trajectory (which
is either $k$ or $k+1$), and $x_{1:\nu}=(x_{1},...,x_{\nu})$ represents
the sequence of target states, with $\nu$ being the length of the
trajectory.

More specifically, a two-step trajectory $X_{k+1}\in T_{(k+1)}$ has
the form\begin{subnumcases}
{X_{k+1}=\label{eq:traj_state_def}}
\left(k,x_{1}\right) & $X_{k+1}\ensuremath{\in}T_{(k+1)}^{1}$ \label{eq:diedState}\\
\left(k+1,x_{1}\right) & $X_{k+1}\ensuremath{\in}T_{(k+1)}^{2}$ \label{eq:bornState}\\
\left(k,x_{1:2}\right) & $X_{k+1}\ensuremath{\in}T_{(k+1)}^{3}.$\label{eq:aliveState}
\end{subnumcases} Line (\ref{eq:diedState}) indicates a trajectory that died in the
time window, i.e., it exists at time step $k$, but it does not at
time step $k+1$. Line (\ref{eq:bornState}) indicates a trajectory
that was born at time step $k+1$. Line (\ref{eq:aliveState}) indicates
a trajectory that exists at time step $k$ and $k+1$ (with two states
$x_{1}$ and $x_{2}$, $x_{1:2}\in\mathbb{R}^{2n_{x}}$).

A set of two-step trajectories is denoted by $\mathbf{X}_{k+1}\in\mathcal{F}\left(T_{(k+1)}\right)$,
where $\mathcal{F}\left(T_{(k+1)}\right)$ denotes the set of all
finite subsets of $T_{(k+1)}$. All mathematical formalities on sets
of trajectories can be found in \cite{GarciaFernandez2020}.

\subsubsection{Single-trajectory measurement and set of trajectory measurements
over $k$ and $k+1$\label{subsec:Single-trajectory-measurement-an}}

We denote a single $n_{z}$-dimensional measurement as $z\in\mathbb{R}^{n_{z}}$,
and the set of measurements at time $k$ as $\mathbf{z}_{k}\in\mathcal{F}(\mathbb{R}^{n_{z}})$.
Analogously to (\ref{eq:tracklets:traj_space}), we define the space
of trajectory measurements $M_{(k+1)}$ in the time window from $k$
to $k+1$ as the disjoint union of three spaces: measurement space
with a detection at time step $k$ but not at time step $k+1$, measurement
space with a detection only at time step $k+1$, and measurement space
with a detection at both time steps $k$ and $k+1$. That is, mathematically,
we can write
\begin{align}
M_{(k+1)} & =M_{(k+1)}^{1}\uplus M_{(k+1)}^{2}\uplus M_{(k+1)}^{3}\nonumber \\
 & =\left[\{k\}\times\mathbb{R}^{n_{z}}\right]\uplus\left[\{k+1\}\times\mathbb{R}^{n_{z}}\right]\uplus\left[\{k\}\times\mathbb{R}^{2n_{z}}\right].\label{eq:meas_space}
\end{align}
Given a trajectory measurement $\left(t,z_{1:m}\right)\in M_{(k+1)}$,
$t$ represents the initial time step of the trajectory measurement,
and $z_{1:\iota}=(z_{1},...,z_{\iota})$ represents the sequence of
$\iota$ measurements.

Therefore, a trajectory measurement $Z_{k+1}$ in the interval from
$k$ and to $k+1$ has the form\begin{subnumcases}
{Z_{k+1}=\label{eq:def_tracklet}}
\left(k,z_{1}\right) & $Z_{k+1}\ensuremath{\in}M_{(k+1)}^{1}$ \label{eq:diedTracklet}\\
\left(k+1,z_{1}\right) & $Z_{k+1}\ensuremath{\in}M_{(k+1)}^{2}$ \label{eq:bornTracklet}\\
\left(k,z_{1:2}\right) & $Z_{k+1}\ensuremath{\in}M_{(k+1)}^{3},$\label{eq:aliveTracklet}
\end{subnumcases}where $z_{1}\in\mathbb{R}^{n_{z}}$ is an $n_{z}$-dimensional measurement,
and $z_{1:2}$ expresses the measurement at both time steps. We call
a trajectory measurement a full (trajectory) measurement if it contains
two measurements at two time steps. On the other hand, the trajectory
measurement is defined as partial if it only contains a measurement
at one time step. A set of trajectory measurements is denoted by $\mathbf{Z}_{k+1}\in\mathcal{F}\left(M_{(k+1)}\right)$.

\subsubsection{Trajectory integrals\label{subsec:Trajectory-integrals}}

Given a trajectory at the current time window $X_{k+1}=(t,x_{1:\nu})\in T_{(k+1)}$,
the variable $\left(t,\nu\right)$ belongs to the set $I_{(k:k+1)}=\left\{ \left(t,\nu\right):t\in\left\{ k,k+1\right\} \text{ and }1\leq\nu\leq k-t+1\right\} $.
The integral of a real-valued function $\pi(\cdot)$ on the single-trajectory
space $T_{(k+1)}$ is \cite{GarciaFernandez2020}
\begin{align}
\ensuremath{\int\pi\left(X\right)dX} & =\sum_{(t,\nu)\in I_{(k:k+1)}}\int\pi\left(t,x_{1:\nu}\right)dx_{1:\nu}\nonumber \\
 & =\int\pi\left(k,x_{1:2}\right)dx_{1:2}+\int\pi\left(k,x_{1}\right)dx_{1}\nonumber \\
 & \quad+\int\pi\left(k+1,x_{1}\right)dx_{1}.\label{eq:tracklets:single_target_integral}
\end{align}
This integral spans across all potential start times, durations, and
states of the two-step trajectory.

Given a real-valued function $\pi(\cdot)$ on the space $\mathcal{F}\left(T_{(k+1)}\right)$
of sets of two-step trajectories, its set integral is
\begin{equation}
\ensuremath{\int\pi\left(\mathbf{X}\right)\delta\mathbf{X}=\sum_{n=0}^{\infty}\frac{1}{n!}\int\pi\left(\{X_{1},...,X_{n}\}\right)dX_{1:n}},\label{eq:tracklets:set_trajectories_integral}
\end{equation}
where $X_{1:n}=\left(X_{1},\dots,X_{n}\right)$.

\subsection{Dynamic model\label{subsec:tracklets:trans_model_gen}}

We consider the standard dynamic model for sets of targets \cite{Mahler14}.
That is, given a set of targets $\mathbf{x}_{k}$, each target $x_{k}\in\mathbf{x}_{k}$
at time step $k$ survives to the time step $k+1$ with a probability
of survival $p^{S}(x_{k})$, according to a Markov transition density
$g(x_{k+1}|x_{k})$. Alternatively, the target dies with probability
$1-p^{S}(x_{k})$. At each time step, targets are born independently
of existing targets according to a PPP with intensity $\lambda_{k}^{B}(x_{k})$.
Given this transition model for sets of targets, we can write it in
terms of the trajectories.

\subsubsection{Transition model for trajectories in $k$ and $k+1$\label{subsec:tracklets:Transition-model-for-trajectories}}

This section explains the transition model of the set of targets at
time $k$ when we keep the trajectory information of the target at
time steps $k$ and $k+1$. We keep the information on all trajectories
\cite{Granstroem2019}, as this will be required for the update step.

Given the set of target states $\mathbf{x}_{k}$ at time step $k$,
each target state $x\in\mathbf{x}_{k}$ evolves with probability $1$
into a trajectory $Y=(t,y_{1:\nu})$, defined between time steps $k$
and $k+1$. In this case, the transition density is \cite{Granstroem2019}\begin{subnumcases}
{g_{k+1}(Y|x)=\label{eq:transition_density-GM}}
\delta_{k}[t]\delta_{x}(y_{1})p^{S}(x)g(y_{2}|x) & $Y=(k,y_{1:2})$\label{eq:aliveTrans}\\
\delta_{k}[t]\delta_{x}(y_{1})\left(1-p^{S}(x)\right) & $Y=(k,y_{1})$ \label{eq:dyingTrans}\\
0 & otherwise.
\end{subnumcases}where $\delta_{k}[\cdot]$ is a Kronecker delta located at $k$, and
$\delta_{x}(\cdot)$ is a Dirac delta located at $x$. Line (\ref{eq:aliveTrans})
corresponds to the case where the trajectory is present at time steps
$k$ and $k+1$, while in (\ref{eq:dyingTrans}) the trajectory is
only present at time step $k$.

Note that the probability that each target at time step $k$ is kept
in the set of trajectories from time step $k$ and $k+1$ is one,
as this set includes all trajectories at time steps $k$ and $k+1$.
This also happens in the dynamic model for tracking the set of all
trajectories if we keep information on the set of all trajectories
up to the current time step \cite{Xia2020}.

\subsubsection{Birth model\label{subsec:tracklets:Birth-model}}

The multi-trajectory state $\mathbf{X}_{k+1}$ is the union of the
surviving trajectories and new born trajectories. As explained before,
new targets are born independently at each time step following a PPP
with intensity $\lambda_{k+1}^{B}(\cdot$). This implies that the
intensity for new born trajectories is \cite{GarciaFernandez2020a}
\begin{equation}
\lambda_{k+1}^{B}(X)=\begin{cases}
\lambda_{k+1}^{B}(x_{1}) & X=(k+1,x_{1})\\
0 & \text{otherwise.}
\end{cases}\label{eq:B_rate_trj}
\end{equation}

\subsection{Trajectory-measurement model\label{subsec:tracklets:meas_model}}

This section explains the model of the trajectory measurements we
consider. Each two-step trajectory $X\in T_{(k+1)}^{\tau}$, $\tau\in\{1,2,3\}$,
is detected with probability $p^{D}(X_{k+1})$ and generates one trajectory
measurement $Z\in M_{(k+1)}^{\mu}$, $\mu\in\{1,2,3\}$, with density

\begin{flushleft}
$l\left(Z|X\right)$
\end{flushleft}
\begin{subnumcases}
{=\label{eq:meas_model_gen}}
\gamma h_{1,3}(z_{1}|x_{1:2}) & $Z=(k,z_{1}),X=(k,x_{1:2})$\label{eq:l_case2}\\
\gamma h_{2,3}(z_{1}|x_{1:2}) & $Z=(k+1,z_{1}),X=(k,x_{1:2})$\label{eq:l_case3}\\
\widetilde{p}^{D}h_{3,3}(z_{1:2}|x_{1:2}) & $Z=(k,z_{1:2}),X=(k,x_{1:2})$\label{eq:l_case1}\\
h_{1,1}(z_{1}|x_{1}) & $Z=(k,z_{1}),X=(k,x_{1})$\label{eq:l_case5}\\
h_{2,2}(z_{1}|x_{1}) & $Z=(k+1,z_{1}),X=(k+1,x_{1})$\label{eq:l_case9}\\
0 & otherwise,
\end{subnumcases}where $\widetilde{p}^{D}$ is the probability of receiving a full
trajectory measurement conditioned on the existence of a trajectory
$X=(k,x_{1:2})$, and $h_{\mu,\tau}(\cdot|\cdot)$ are densities on
the single-trajectory measurement space, conditioned on the target
state. Finally, we set $\gamma=0.5\left(1-\widetilde{p}^{D}\right)$
such that the integral of $l(Z|X)$ (over the single-trajectory space),
given $X=(k,x_{1:2})$, is unity and the detection of either end of
the trajectory has equal probability. Alternatively, the trajectory
is not detected with probability $1-p^{D}(X)$. 

Clutter is distributed according to a PPP with intensity $\lambda^{C}(\cdot)$
on the trajectory-measurement space. More information on PPPs for
sets of trajectories can be found in \cite{GarciaFernandez2019a}.

\subsection{PMBM posterior and predicted densities\label{subsec:PMBM-solution-overview}}

Given the dynamic and measurement models, our objective is to compute
the posterior density $f_{k+1|k+1}(\cdot)$ over the set of targets
given the sequence of trajectory measurements up to the time step
$k+1$. For these models, both the posterior and predicted densities
are PMBMs \cite{GarciaFernandez2020a,Granstroem2019}. We proceed
to give an overview of the PMBM posterior and predicted densities
for both targets and trajectories.

\subsubsection{PMBM density on targets\label{subsec:PMBM-posterior-density}}

The PMBM posterior and predicted densities over the set of target
states $\mathbf{x}_{k'}$ at time step $k'$, with $k'\in\{k,k+1\}$,
are expressed as \cite{Garcia-Fernandez18}
\begin{align}
f_{k+1|k'}\left(\mathbf{x}_{k'}\right)= & \sum_{\uplus_{l=1}^{n_{k+1|k'}}\mathbf{x}^{l}\uplus\mathbf{y}=\mathbf{x}_{k'}}f_{k+1|k'}^{p}\left(\mathbf{y}\right)f_{k+1|k}^{mbm}(\mathbf{x}),\label{eq:PMBM_on_states}
\end{align}
where the sum goes over all mutually disjoint sets $\mathbf{y}$ and
$\mathbf{x}$, such that their union is $\mathbf{x}_{k'}$. The two
densities in (\ref{eq:PMBM_on_states}) are explained next.

The PPP density represents the targets that exist at the current time
instant, but have not yet been detected. Its density is
\begin{equation}
f_{k+1|k'}^{p}(\mathbf{x})=e^{-\int\lambda_{k+1|k'}(x)dx}\prod_{x\in\mathbf{x}}\lambda_{k+1|k'}(x),\label{eq:PPP-3}
\end{equation}
where $\lambda_{k+1|k'}(\cdot)$ is the intensity. In the PPP, the
cardinality is Poisson distributed and targets are independent, and
identically distributed. The MBM part represents the potential targets
that have been detected at some point up to the current time step,
and it can be described as \cite{Williams15a}:
\begin{equation}
f_{k+1|k'}^{mbm}(\mathbf{x})=\sum_{a\in\mathcal{A}_{k+1|k'}}w_{_{k+1|k'}}^{a}\sum_{\uplus_{j=1}^{n_{k+1|k'}}\mathbf{x}^{j}=\mathbf{x}}\thinspace\prod_{i=1}^{n_{k+1|k'}}f_{k+1|k'}^{i,a^{i}}(\mathbf{x}^{i}),\label{eq:MBM-3}
\end{equation}
where $i$ is the index over the Bernoulli components, $a=(a^{1},\dots,a^{n_{k+1|k'}})$
represents a specific global data association hypothesis, $a^{i}\in\{1,\dots,h_{k+1|k'}^{i}\}$
is an index over the $h_{k+1|k'}^{i}$ single target hypotheses for
the $i$-th potential target, $\mathcal{A}_{k+1|k'}$ is the set of
global hypotheses, and $n_{k+1|k'}$ is the number of potentially
detected targets. Each global hypothesis is associated to a weight
$w_{_{k+1|k'}}^{a}$ satisfying $\sum_{a\in\mathcal{A}_{k+1|k'}}w_{_{k+1|k'}}^{a}=1$.
The Bernoulli density corresponding to the $i$-th potential target,
$i\in\{1,\dots n_{k+1|k'}\}$ and the $a_{i}$ single target hypothesis
density $f_{k+1|k'}^{i,a^{i}}(\cdot)$ can be expressed as\begin{subnumcases}
{f_{k+1|k'}^{i,a^{i}}\left(\mathbf{x}\right)=\label{eq:Bernoulli_density_filter}}
1-r_{k+1|k'}^{i,a^{i}} & $\mathbf{x}=\emptyset$\\
r_{k+1|k'}^{i,a^{i}}p_{k+1|k'}^{i,a^{i}}(x) & $\mathbf{x}=\left\{ x\right\}$\\
0 & otherwise,
\end{subnumcases}where $r_{k+1|k'}^{i,a^{i}}\in[0,1]$ is the probability of existence
and $p_{k+1|k'}^{i,a^{i}}(\cdot)$ is the single-target density given
that it exists.

\subsubsection{PMBM on sets of trajectories from $k$ to $k+1$\label{subsec:PMBM-on-trajectories}}

Following \cite{GarciaFernandez2020a,Granstroem2019}, we extend the
PMBM filtering/predicted density over the set of two-step trajectories.
Given a PMBM density on the set of target states at the time step
$k$ of the form (\ref{eq:PMBM_on_states}), the transition and measurement
models for trajectories indicated in Sections \ref{subsec:tracklets:Transition-model-for-trajectories}
and \ref{subsec:tracklets:meas_model}, the predicted and updated
densities on the set of two-step trajectories $\mathbf{X}_{k+1}$
from time step $k$ to $k+1$ are also PMBMs of the form
\begin{align}
f_{k+1|k'}\left(\mathbf{X}_{k'}\right)= & \sum_{\uplus_{l=1}^{n_{k+1|k'}}\mathbf{X}^{l}\uplus\mathbf{Y}=\mathbf{X}_{k'}}f_{k+1|k'}^{p}\left(\mathbf{Y}\right)f_{k+1|k'}^{mbm}(\mathbf{X}),\label{eq:PMBM_on_trj}
\end{align}
where $n_{k+1|k}=n_{k|k}$ and $f_{k+1|k'}^{p}(\cdot)$ and $f_{k+1|k'}^{mbm}(\cdot)$
are the PPP and MBM densities for the two-time step trajectories.
This result is obtained as a consequence of the prediction and update
steps of the trajectory PMBM filter (PMBM filter on the sets of trajectories)
\cite{Granstroem2018}. It should be noted that when we receive trajectory
measurements, the local and global hypotheses are defined as for standard
measurements, with the only difference that standard point target
measurements are now two-step trajectories. To continue with the PMBM
filtering recursion on the set of targets, the density (\ref{eq:PMBM_on_trj})
is then marginalised to obtain a density of the form (\ref{eq:PMBM_on_states})
on the target states $\mathbf{x}_{k+1}$ at time $k+1$.

\section{TM-PMBM filtering recursion\label{sec:tracklets:filtering_recursion}}

This section presents the filtering recursion of the TM-PMBM filter,
whose diagram is provided in Fig. \ref{fig:tracklets:filter-schematic}.
At each time step $k$, the PMBM density on the set of targets $\mathbf{x}_{k}$
is of the form (\ref{eq:PMBM_on_states}). In Section \ref{subsec:tracklets:Prediction},
we perform the prediction step based on the transition density (\ref{eq:transition_density-GM})
to obtain the predicted PMBM density (\ref{eq:PMBM_on_trj}) on the
set of two-step trajectories $\mathbf{X}_{k+1}$ in the time window
$k+1$, defined by the time steps $k$ and $k+1$. In Section \ref{subsec:tracklets:Update},
we update the density on two-step trajectories based on the set of
trajectory measurements $\mathbf{Z}_{k+1}$ in the time window $k+1.$
In Section \ref{subsec:Marginalisation}, the density (\ref{eq:PMBM_on_trj})
is then marginalised to obtain a density of the form (\ref{eq:PMBM_on_states})
on the target states $\mathbf{x}_{k+1}$ at time $k+1$.

In this section, we use the following notation. Given two real-valued
functions $a(\cdot)$ and $b(\cdot)$ on either the single-target
space or the single-trajectory space, their inner product is denoted
by \cite{GarciaFernandez2021,GarciaFernandez2020a}
\begin{equation}
\left\langle a,b\right\rangle =\int a(X)b(X)\,dX.\label{eq:tracklets:inner_prod}
\end{equation}

\subsection{Prediction\label{subsec:tracklets:Prediction}}

The predicted PMBM density over the set of two-step trajectories is
derived from the posterior PMBM density after the marginalisation
described in Section \ref{subsec:Marginalisation}. For clarity, we
use the superscript $M$ to denote densities and parameters that define
the marginalised posterior at time step $k$.

With the transition model in (\ref{eq:transition_density-GM}), and
applying the prediction step of the TPMBM filter for the set of all
trajectories \cite{Granstroem2018}, the predicted PMBM density is
of the form (\ref{eq:PMBM_on_trj}) with $n_{k+1|k}=n_{k|k}$.
\begin{lem}
[TM-PMBM prediction]Given the PMBM filtering density on the set of
target states at time step $k$ of the form (\ref{eq:PMBM_on_states}),
the predicted density at time $k+1$ is a PMBM of the form (\ref{eq:PMBM_on_trj}),
with $n_{k+1|k}=n_{k|k}$ and Poisson intensity \label{lem:TM-PMBM_prediction}
\begin{align}
\lambda_{k+1|k}(X_{k+1}) & =\lambda_{k+1}^{B}\left(X_{k+1}\right)+\left\langle \lambda_{k|k}^{M},g_{k+1}(X_{k+1}|\cdot)\right\rangle .\label{eq:PPP_gen_pred}
\end{align}
Each Bernoulli component $f_{k+1|k}^{i,a^{i}}(\cdot)$, $i\in\{1,\dots,n_{k+1|k}\}$,
$a^{i}\in\{1,\dots,h_{k+1|k}^{i}\}$, in the MBM part is defined by
\begin{align}
r_{k+1|k}^{i,a^{i}} & =r_{k|k}^{i,a^{i},M}\label{eq:r_gen_pred}\\
p_{k+1|k}^{i,a^{i}}(X_{k+1}) & =\left\langle p_{k|k}^{i,a^{i},M},g_{k+1}(X_{k+1}|\cdot)\right\rangle .\label{eq:single_state_density_gen_pred}
\end{align}
\end{lem}

\subsection{Update\label{subsec:tracklets:Update}}

Given the trajectory measurement model in Section \ref{subsec:tracklets:meas_model},
the TM-PMBM filter update is provided in the following lemma.
\begin{lem}
[TM-PMBM update]Given a set of trajectory measurements $\mathbf{Z}_{k+1}=\{Z_{k+1}^{1},\dots,Z_{k+1}^{m_{k+1}}\}$
in the time window between time steps $k$ and $k+1$, and a PMBM
predicted density on the set of two-step trajectories of the form
(\ref{eq:PMBM_on_trj}), the updated distribution is exactly a PMBM
density of the form (\ref{eq:PMBM_on_trj}) with $n_{k+1|k+1}=n_{k+1|k}+m_{k+1}$
and Poisson intensity \label{lem:TM-PMBM update}
\begin{equation}
\lambda_{k+1|k+1}(X)=\left(1-p^{D}(X)\right)\lambda_{k+1|k}(X).\label{eq:PPP_gen_upd}
\end{equation}
For each Bernoulli component in $f_{k+1|k}^{i,a^{i}}(\cdot)$ , $i\in\{1,\dots,n_{k+1|k}\}$,
$a^{i}\in\{1,\dots,h_{k+1|k}^{i}\}$, there is a missed detection
hypothesis with the following parameters
\begin{align}
w_{k+1|k+1}^{i,a^{i}} & =1-r_{k+1|k}^{i,a^{i}}\left\langle p_{k+1|k}^{i,a^{i}},p^{D}\right\rangle \label{eq:w_gen_upd_mis-1}\\
r_{k+1|k+1}^{i,a^{i}} & =\frac{r_{k+1|k}^{i,a^{i}}\left\langle p_{k+1|k}^{i,a^{i}},1-p^{D}\right\rangle }{1-r_{k+1|k}^{i,a^{i}}\left\langle p_{k+1|k}^{i,a^{i}},p^{D}\right\rangle }\label{eq:r_gen_update_mis-1}\\
p_{k+1|k+1}^{i,a^{i}}(X) & =\frac{\left(1-p^{D}(X)\right)p_{k+1|k}^{i,a^{i}}(X)}{\left\langle p_{k+1|k}^{i,a^{i}},1-p^{D}\right\rangle }.\label{eq:single_state_density_gen_upd_mis-1}
\end{align}
The hypothesis for the existing Bernoulli component $f_{k+1|k}^{i,\widetilde{a}^{i}}(\cdot)$,
$i\in\{1,\dots,n_{k+1|k}\}$, $\widetilde{a}^{i}\in\{1,\dots,h_{k+1|k}^{i}\}$,
and trajectory measurement $Z_{k+1}^{j}$ has index $a^{i}=\widetilde{a}^{i}+h_{k+1|k}^{i}$
and parameters
\begin{align}
r_{k+1|k+1}^{i,a^{i}} & =1\label{eq:existence_prob_gen_upd-1}\\
w_{k+1|k+1}^{i,a^{i}} & =r_{k+1|k}^{i,\widetilde{a}^{i}}\left\langle p_{k+1|k}^{i,\widetilde{a}^{i}},l\left(Z_{k+1}^{j}|\cdot\right)p^{D}\right\rangle \label{eq:w_gen_upd_det-1}\\
p_{k+1|k+1}^{i,a^{i}}(X) & =\frac{l\left(Z_{k+1}^{j}|X\right)p^{D}(X)p_{k+1|k}^{i,\widetilde{a}^{i}}(X)}{\left\langle p_{k+1|k}^{i,\widetilde{a}^{i}},l\left(Z_{k+1}^{j}|\cdot\right)p^{D}(\cdot)\right\rangle }.\label{eq:single_state_density_gen_upd_det-1}
\end{align}
For a new Bernoulli component $i=n_{k+1|k}+j$, $j\in\{1,\dots,m_{k+1}\}$,
initiated by the trajectory measurement $Z_{k+1}^{j}$, there are
two local hypotheses: missed detection with parameters $w_{k+1|k+1}^{i,1}=1$,
$r_{k+1|k+1}=0$, and detection with parameters
\begin{align}
w_{k+1|k+1}^{i,2} & =\lambda^{C}(Z_{k+1}^{j})+\left\langle \lambda_{k+1|k},l\left(Z_{k+1}^{j}|\cdot\right)p^{D}(\cdot)\right\rangle \label{eq:tracklets:w_gen_upd_new}\\
r_{k+1|k+1}^{i,2} & =\frac{\left\langle \lambda_{k+1|k},l\left(Z_{k+1}^{j}|\cdot\right)p^{D}(\cdot)\right\rangle }{\lambda^{C}(Z_{k+1}^{j})+\left\langle \lambda_{k+1|k},l\left(Z_{k+1}^{j}|\cdot\right)p^{D}(\cdot)\right\rangle }\label{eq:r_gen_upd_new-1}\\
p_{k+1|k+1}^{i,2}(X) & =\frac{l\left(Z_{k+1}^{j}|X\right)p^{D}(X)\lambda_{k+1|k}(X)}{\left\langle \lambda_{k+1|k},l\left(Z_{k+1}^{j}|\cdot\right)p^{D}(\cdot)\right\rangle }.\label{eq:single_state_density_gen_upd_new-1}
\end{align}
\end{lem}
This update is a direct result of the update steps of the PMBM filter
\cite{Williams15a} and the TPMBM filter for point targets \cite{Granstroem2019}.
It should be noted that the TM-PMBM update step is applied on the
density on the sets of trajectories between time steps $k$ and $k+1$
because the trajectory measurements contain information on the trajectories
on these two time steps. This is also required to compute the dot
products in this lemma.

\subsection{Marginalisation\label{subsec:Marginalisation}}

In this paper, once we have performed the update in Lemma \ref{lem:TM-PMBM update},
we are only interested in keeping the information on the set of targets
at the end of the time window, i.e., at time step $k+1$. To do so,
we perform a marginalisation step after we have performed the update,
integrating out the trajectory states at time step $k$ while retaining
the information corresponding to time step $k+1$. The marginalisation
theorem for general densities on sets of trajectories is provided
in Theorem 11 in \cite{GarciaFernandez2020}. The marginalisation
theorem applied to PMBMs on sets of trajectories is provided in \cite{Granstroem2019}.
The result is another PMBM on the set of targets at time step $k+1$.

The marginalisation step can be carried out in an exact manner by
choosing a suitable multi-target transition model (distinct from the
one used in the prediction step) and performing a PMBM prediction
step. We proceed to explain the multi-target transition model to perform
the marginalisation, which includes a probability of survival and
a single-target transition density.

We consider a two-step trajectory $X_{k+1}=(t,x_{1:\nu})\in T_{(k+1)}$.
The probability of survival at the marginalisation stage is \begin{subnumcases}
{p_{k+1}^{S,M}\left(X_{k+1}\right)=\label{eq:survival_marginalisation}}
0 & $X_{k+1}\ensuremath{\in}T_{(k+1)}^{1}$\\
1 & $X_{k+1}\ensuremath{\in}T_{(k+1)}^{2}$\\
1 & $X_{k+1}\ensuremath{\in}T_{(k+1)}^{3}$.
\end{subnumcases}That is, two-step trajectories that belong to $T_{(k+1)}^{1}$, are
those that do not exist at time step $k$ so they are not included
in the set of targets at time step $k+1$. On the contrary, two-step
trajectories that belong to $T_{(k+1)}^{2}$ and $T_{(k+1)}^{3}$
have a target state at time step $k+1$ so their probability of survival
is one. Then, we apply the single-target transition density to each
surviving target\begin{subnumcases}
{g_{k+1}^{M}\left(x_{k+1}|X\right)=\label{eq:marginal_transDensity}}
\delta_{x_{2}}\left(x_{k+1}\right) & $X=\left(k,x_{1:2}\right)$\\
\delta_{x_{1}}\left(x_{k+1}\right) & $X=\left(k+1,x_{1}\right)$\\
0 & otherwise.
\end{subnumcases}where $x_{k+1}$ is a target state. That is, this single-target transition
density simply keeps the target state at time step $k+1$ of the trajectory,
which corresponds to $x_{2}$ if $X=(k,x_{1:2})$ and to $x_{1}$
if $X=(k+1,x_{1})$. When we apply a prediction step using this transition
model to the updated PMBM on the set of trajectories, the result is
a PMBM on the set of targets at time step $k+1$ that discards trajectory
information at time step $k$.
\begin{lem}
[TM-PMBM marginalisation]Given the updated PMBM density on the set
of two-step trajectories between time steps $k$ and $k+1$ of the
form (\ref{eq:PMBM_on_trj}), the PMBM density over the set of target
states is of the form (\ref{eq:PMBM_on_states}) \cite{Granstroem2019},
with PPP intensity \label{lem:TM-PMBM marginalisation}
\begin{align}
\lambda_{k+1|k+1}^{M}(x_{k+1}) & =\left\langle \lambda_{k+1|k+1},g_{k+1}^{M}(x_{k+1}|\cdot)p_{k+1}^{S,M}\left(\cdot\right)\right\rangle ,\label{eq:PPP_gen_marg}
\end{align}
and Bernoulli components $f_{k+1|k+1}^{i,a^{i},M}(\cdot)$, $i\in\{1,\dots,n_{k+1|k+1}\}$,
$a^{i}\in\{1,\dots,h_{k+1|k+1}^{i}\}$, with probability of existence
\begin{align}
r_{k+1|k+1}^{i,a^{i},M} & =r_{k+1|k+1}^{i,a^{i}}\left\langle p_{k+1|k+1}^{i,a^{i}},p_{k+1}^{S,M}\right\rangle \label{eq:r_gen_marg}
\end{align}
and single-target density
\begin{align}
p_{k+1|k+1}^{i,a^{i},M}(x_{k+1}) & =\frac{\left\langle p_{k|k}^{i,a^{i}},g_{k+1}^{M}(x_{k+1}|\cdot)p_{k+1}^{S,M}(\cdot)\right\rangle }{\left\langle p_{k|k}^{i,a^{i}},p_{k+1}^{S,M}\right\rangle }.\label{eq:single_state_density_gen_marg}
\end{align}
\end{lem}

\section{Gaussian TM-PMBM filter\label{sec:tracklets:Gaussian-TM-PMBM-filter}}

In this section, we present the Gaussian implementation of the TM-PMBM
filter. Section \ref{subsec:Preliminaries_GM} outlines the assumptions
underlying the Gaussian implementation. Section \ref{subsec:Prediction_GM}
details the prediction step for both the PPP and Bernoulli components.
Finally, Section \ref{subsec:Update_GM} provides the equations for
the update step, addressing detection, missed detection, and new-birth
hypotheses. The pseudocode of the TM-PMBM filter is given in Algorithm
\ref{alg:TM_PMBM_pseoudocode}.

\subsection{Preliminaries\label{subsec:Preliminaries_GM}}

From \cite{GarciaFernandez2020a}, we use the notation\begin{subnumcases}
{\mathcal{N}(t,x_{1:\nu};k,\overline{x},P)=\label{eq:Gaussian_traj_notation-1}}
\mathcal{N}(x_{1:\nu};\overline{x},P) & $t=k,\,\nu=\iota$\\
0 & otherwise
\end{subnumcases}to represent a trajectory Gaussian density with start time $k$ and
number of states $\iota$, where $\iota=\dim(\overline{x})/n_{x}$.
The mean $\overline{x}\in\mathbb{R}^{n_{x}\iota}$ and covariance
matrix $P\in\mathbb{R}^{n_{x}\iota\times n_{x}\iota}$ are defined
according to the length of the trajectory. The notation (\ref{eq:Gaussian_traj_notation-1})
can also be applied to trajectory measurements defined in the space
$M_{(k+1)}$.

For the Gaussian implementation we make the following assumptions:
\begin{itemize}
\item The probability of survival is constant $p^{S}(x)=p^{S}$.
\item The probability of detection is constant $p^{D}(X_{k+1})=p^{D}.$
\item The transition density is Gaussian $g(\cdot|x)=\mathcal{N}(\cdot;Fx,Q)$,
where $F,Q\in\mathbb{R}^{n_{x}\times n_{x}}$.
\item $H$ is an $n_{z}\times n_{x}$ matrix defining the measurement model
for a single target state and point detection.
\item $R$ is an $n_{x}\times n_{x}$ covariance matrix of the measurement
noise for a single target state and point detection.
\item A two-step trajectory $X$ generates a trajectory measurement $Z$
according to (\ref{eq:meas_model_gen}), where
\[
h_{\mu,\tau}(z_{1:\iota}|x_{1:\nu})=\mathcal{N}\left(z_{1:\iota};H_{\mu,\tau}x_{1:\nu},R_{\iota}\right),
\]
with $R_{\iota}=I_{\iota}\otimes R$, $\iota=1$ for $\mu\in{1,2}$
and $\iota=2$ for $\mu=3$, and $H_{\mu,\tau}$ is\begin{subnumcases}
{H_{\mu,\tau}=\label{eq:H_cases}}
[1,0]\otimes H & $\mu=1,\tau=3$\\{}
[0,1]\otimes H & $\mu=2,\tau=3$\\
I_{2}\otimes H & $\mu=3,\tau=3$\\
H & $\mu=1,\tau=1$\\
H & $\mu=2,\tau=2$\\
0 & otherwise.
\end{subnumcases}
\item The intensity of new born trajectories is
\begin{equation}
\lambda_{k+1}^{B}(X_{k+1})=\sum_{q_{b}=1}^{n_{k+1}^{b}}w_{k+1}^{b,q_{b}}\mathcal{N}(X_{k+1};k+1,\overline{x}_{k+1}^{b,q_{b}},P_{k+1}^{b,q_{b}}),\label{eq:PHD_at_k+1}
\end{equation}
where $n_{k+1}^{b}$ is the number of components, $w_{k+1}^{b,q_{b}}$
is the weight of the $q_{b}$-th component, $\overline{x}_{k+1}^{b,q_{b}}\in\mathbb{R}^{n_{x}}$
its mean and $P_{k+1}^{b,q_{b}}\in\mathbb{R}^{n_{x}\times n_{x}}$
its covariance matrix. That is, new born trajectories have length
1 with probability one.
\end{itemize}
Given these assumptions, the PPP intensity on a target state $x$
is a Gaussian mixture 
\begin{equation}
\lambda_{k|k}^{M}(x)=\sum_{q=1}^{n_{k+1|k'}^{p}}w_{k|k}^{p,q}\mathcal{N}(x;\overline{x}_{k|k}^{p,q,M},P_{k|k}^{p,q,M}),\label{eq:PHD_Gaus_gen_tar}
\end{equation}
where $w_{k|k}^{p,q}>0$, $\overline{x}_{k|k}^{p,q,M}\in\mathbb{R}^{n_{x}}$
and $P_{k|k}^{p,q,M}\in\mathbb{R}^{n_{x}\times n_{x}}$. The PPP intensity
on a two-step trajectory $X$ at time step $k+1,$ given the measurements
up to $k'={k,k+1}$, is a Gaussian mixture 
\begin{equation}
\lambda_{k+1|k'}(X)=\sum_{q=1}^{n_{k+1|k'}^{p}}w_{k+1|k'}^{p,q}\mathcal{N}(X;t_{k+1|k'}^{p,q},\overline{x}_{k+1|k'}^{p,q},P_{k+1|k'}^{p,q}),\label{eq:PHD_Gaus_gen_trj}
\end{equation}
where $n_{k+1|k'}^{p}$ is the number of components, $t_{k+1|k'}^{p,q}\in\{k,k+1\}$,
$w_{k+1|k'}^{p,q}$ is the weight of the $q$-th component, $\overline{x}_{k+1|k'}^{p,q}\in\mathbb{R}^{2n_{x}}$
and $P_{k+1|k'}^{p,q}\in\mathbb{R}^{2n_{x}\times2n_{x}}$.

Furthermore, the Bernoulli component $f_{k+1|k'}^{i,a^{i}}(\cdot)$,
$i\in\{1,\dots,n_{k+1|k'}\}$, $a^{i}\in\{1,\dots,h_{k+1|k'}^{i}\}$,
on a target state $x$ has a single-target density 
\begin{equation}
p_{k|k}^{i,a^{i},M}(x)=\mathcal{N}(x;\overline{x}_{k|k}^{i,a^{i},M},P_{k|k}^{i,a^{i},M}).\label{eq:Gaus_single_tar_density}
\end{equation}
The correspondent Bernoulli component on a two-step trajectory $X$
has a single-trajectory density
\begin{equation}
p_{k+1|k'}^{i,a^{i}}(X)=\sum_{l=1}^{2}\beta_{k+1|k'}^{i,a^{i}}(l)\mathcal{N}(X;k,\overline{x}_{k+1|k'}^{i,a^{i}}(l),P_{k+1|k'}^{i,a^{i}}(l)),\label{eq:Gaus_single_trj_density}
\end{equation}
where $\beta_{k+1|k'}^{i,a^{i}}(l)$ represents the probability that
the corresponding trajectory terminates at time step $k$ when $l=1$,
or survives until time step $k+1$ when $l=2$, and $\overline{x}_{k+1|k+1}^{i,a^{i}}(l)\in\mathbb{R}^{ln_{x}}$
and $P_{k+1|k+1}^{i,a^{i}}(l)\in\mathbb{R}^{ln_{x}\times ln_{x}}$.
We denote the components of the mean vector and covariance matrix
as
\begin{align}
\overline{x}_{k+1|k'}^{i,a^{i}}(2) & =\left[\begin{array}{c}
\overline{x}_{k+1|k'}^{i,a^{i},1}(2)\\
\overline{x}_{k+1|k'}^{i,a^{i},2}(2)
\end{array}\right]\label{eq:mean_upd_componets}\\
P_{k+1|k'}^{i,a^{i}}(2) & =\left[\begin{array}{cc}
P_{k+1|k'}^{i,a^{i},1,1}(2) & P_{k+1|k'}^{i,a^{i},1,2}(2)\\
P_{k+1|k'}^{i,a^{i},2,1}(2) & P_{k+1|k'}^{i,a^{i},2,2}(2)
\end{array}\right]\label{eq:covar_upd_components}
\end{align}
for cases with $l=2$ in (\ref{eq:Gaus_single_trj_density}).

\subsection{Prediction\label{subsec:Prediction_GM}}

Based on Lemma \ref{lem:TM-PMBM_prediction}, we compute the prediction
step in the Gaussian implementation. As in the standard TPMBM/TPMB
filter implementations, in the PPP, we only consider alive trajectories
\cite{GarciaFernandez2020a}.
\begin{lem}
[Gaussian TM-PMBM prediction]\label{lem:Gaus_TMPMBM_pred}Assume
the filtering density at time step $k$ is a PMBM on the set of target
states of the form (\ref{eq:PMBM_on_states}), with PPP intensity
of the form (\ref{eq:PHD_Gaus_gen_tar}). According to the transition
density (\ref{eq:transition_density-GM}), the predicted intensity
(\ref{eq:PPP_gen_pred}) is a Gaussian mixture intensity
\begin{align}
\lambda_{k+1|k}(X_{k+1}) & =\underbrace{\sum_{q_{b}=1}^{n_{k+1}^{b}}w_{k+1}^{b,q_{b}}\mathcal{N}(X_{k+1};k+1,\overline{x}_{k+1}^{b,q_{b}},P_{k+1}^{b,q_{b}})}_{\lambda_{k+1}^{B}\left(X_{k+1}\right)}\nonumber \\
 & \quad+\sum_{q_{p}=1}^{n_{k+1|k}^{p}}w_{k+1|k}^{p,q_{p}}\mathcal{N}(X_{k+1};k,\overline{x}_{k+1|k}^{p,q_{p}},P_{k+1|k}^{p,q_{p}}),\label{eq:PPP_GM}
\end{align}
where $n_{k+1|k}^{p}=n_{k}^{b}+n_{k|k}^{p}$, $w_{k+1}^{b,q_{b}}$
and $w_{k+1|k}^{p,q_{p}}$ are the weights of the $q_{b}$-th and
$q_{p}$-th components, with 
\begin{equation}
w_{k+1|k}^{p,q_{p}}=w_{k|k}^{p,q_{p}}p^{S},\label{eq:weight_PPP_pred}
\end{equation}
and $\overline{x}_{k+1|k}^{p,q_{p}}$ and $P_{k+1|k}^{p,q_{p}}$ are
the mean and covariance for surviving trajectories
\begin{align}
\overline{x}_{k+1|k}^{p,q_{p}} & =\left[\begin{array}{c}
\overline{x}_{k|k}^{p,q_{p},M}\\
F\overline{x}_{k|k}^{p,q_{p},M}
\end{array}\right]\label{eq:meanPPP_pred}
\end{align}
\begin{equation}
P_{k+1|k}^{p,q_{p}}=\left[\begin{array}{cc}
P_{k|k}^{p,q_{p},M} & P_{k|k}^{p,q_{p},M}F^{T}\\
FP_{k|k}^{p,q_{p},M} & FP_{k|k}^{p,q_{p},M}F^{T}+Q
\end{array}\right],\label{eq:covPPP_pred}
\end{equation}
while the mean $\overline{x}_{k+1}^{b,q_{b}}$ and covariance $P_{k+1}^{b,q_{b}}$
for new born trajectories are defined in (\ref{eq:PHD_at_k+1}).

Assume the Bernoulli component $f_{k|k}^{i,a^{i}}(\cdot)$, $i\in\{1,\dots,n_{k|k}\}$,
$a^{i}\in\{1,\dots,h_{k|k}^{i}\}$ is defined by $r_{k|k}^{i,a^{i},M}$
and single-target density in the form of (\ref{eq:Gaus_single_tar_density}).
According to (\ref{eq:r_gen_pred}) and (\ref{eq:single_state_density_gen_pred}),
the predicted Bernoulli component $f_{k+1|k}^{i,a^{i}}(\cdot)$ has
single-target density of the form (\ref{eq:Gaus_single_trj_density})
with
\begin{align}
r_{k+1|k}^{i,a^{i}} & =r_{k|k}^{i,a^{i},M}\label{eq:R_Gauss_pred}\\
\beta_{k+1|k}^{i,a^{i}}(l) & =\begin{cases}
1-p^{S} & l=1\\
p^{S} & l=2
\end{cases}\label{eq:beta_pred}\\
\overline{x}_{k+1|k}^{i,a^{i}}(1) & =\overline{x}_{k|k}^{i,a^{i},M}\label{eq:single_mean_pred}\\
P_{k+1|k}^{i,a^{i}}(1) & =P_{k|k}^{i,a^{i},M}\label{eq:single_P_pred}\\
\overline{x}_{k+1|k}^{i,a^{i}}(2) & =\left[\begin{array}{c}
\overline{x}_{k|k}^{i,a^{i},M}\\
F\overline{x}_{k|k}^{i,a^{i},M}
\end{array}\right]\label{eq:mean_pred}\\
P_{k+1|k}^{i,a^{i}}(2) & =\left[\begin{array}{cc}
P_{k|k}^{i,a^{i},M} & P_{k|k}^{i,a^{i},M}F^{T}\\
FP_{k|k}^{i,a^{i},M} & FP_{k|k}^{i,a^{i},M}F^{T}+Q
\end{array}\right].\label{eq:cov_pred}
\end{align}
\end{lem}

\subsection{Update\label{subsec:Update_GM}}

Building on Lemma \ref{lem:TM-PMBM update}, we compute the update
step in the Gaussian implementation.
\begin{lem}
[Gaussian TM-PMBM update]\label{lem:Gaus_TMPMBM_upd}Assume the PMBM
density (\ref{eq:PMBM_on_trj}) on the set of two-step trajectories
$\mathbf{X}_{k+1}$ with $\lambda_{k+1|k}(\cdot)$ of the form (\ref{eq:PPP_GM}),
and Bernoulli components $f_{k+1|k}^{i,a^{i}}(\cdot)$, $i\in\{1,\dots,n_{k+1|k}\}$,
$a^{i}\in\{1,\dots,h_{k+1|k}^{i}\}$, with single trajectory density
$p_{k+1|k}^{i,a^{i}}(\cdot)$ of the form (\ref{eq:Gaus_single_trj_density}).
Then, the updated density with the set of trajectory measurements
$\mathbf{Z}_{k+1}=\{Z_{k+1}^{1},\dots,Z_{k+1}^{m_{k+1}}\}$ is a PMBM
with PPP intensity
\begin{align}
 & \lambda_{k+1|k+1}(X_{k+1})=(1-p^{D})\lambda_{k+1|k}(X_{k+1})\nonumber \\
 & =\underbrace{(1-p^{D})\sum_{q_{b}=1}^{n_{k+1}^{b}}w_{k+1}^{b,q_{b}}\mathcal{N}(X_{k+1};k+1,\overline{x}_{k+1}^{b,q_{b}},P_{k+1}^{b,q_{b}})}_{\text{non-detected new-born trajectories}}\nonumber \\
 & \quad+\underbrace{\sum_{q_{p}=1}^{n_{k+1|k}^{p}}w_{k+1|k+1}^{p,q_{p}}\mathcal{N}(X_{k+1};k,\overline{x}_{k+1|k}^{p,q_{p}},P_{k+1|k}^{p,q_{p}})}_{\text{non-detected surviving trajectories}},\label{eq:PPP_update-1}
\end{align}
where
\begin{equation}
w_{k+1|k+1}^{p,q_{p}}=(1-p^{D})w_{k+1|k}^{p,q_{p}}.\label{eq:weight_PPP_upd}
\end{equation}
The updated single-trajectory densities of the Bernoulli components
$p_{k+1|k+1}^{i,a^{i}}(\cdot)$ have the form (\ref{eq:Gaus_single_trj_density}).
The missed detection hypothesis for the Bernoulli component $f_{k+1|k}^{i,a^{i}}(\cdot)$
has 
\begin{align}
w_{k+1|k+1}^{i,a^{i}} & =1-r_{k+1|k}^{i,a^{i}}p^{D}\label{eq:W_mis_update}\\
r_{k+1|k+1}^{i,a^{i}} & =\frac{r_{k+1|k}^{i,a^{i}}(1-p^{D})}{1-r_{k+1|k}^{i,a^{i}}p^{D}},\label{eq:R_mis_update}
\end{align}
and the two components in the mixture (\ref{eq:Gaus_single_trj_density}),
corresponding to $l=\{1,2\}$, have the following parameters
\begin{align}
\beta_{k+1|k+1}^{i,a^{i}}(l) & =\beta_{k+1|k}^{i,a^{i}}(l)\label{eq:beta_mis_update}\\
\overline{x}_{k+1|k+1}^{i,a^{i}}(l) & =\overline{x}_{k+1|k}^{i,a^{i}}(l)\label{eq:mean_gaus_misD}\\
P_{k+1|k+1}^{i,a^{i}}(l) & =P_{k+1|k}^{i,a^{i}}(l).\label{eq:cov_gaus_missD}
\end{align}
The detection hypothesis for the existing Bernoulli component $f_{k+1|k}^{i,\widetilde{a}^{i}}(\cdot)$
and trajectory measurement $Z^{j}=(t,z_{1:\iota})$ is denoted with
index $a^{i}=\widetilde{a}^{i}+h_{k+1|k}^{i}j$, where $i\in\{1,\dots,n_{k+1|k}\}$,
$\widetilde{a}^{i}\in\{1,\dots,h_{k+1|k}^{i}\}$ and $j\in\left\{ 1,\dots,m_{k}\right\} $.
We proceed to explain the update for the different types of measurements.

If $Z^{j}=(k,z_{1})\in M_{(k+1)}^{1}$, then based on (\ref{eq:l_case2})
and (\ref{eq:l_case5}), both the components $l\in\left\{ 1,2\right\} $
in the mixture (\ref{eq:Gaus_single_trj_density}) are non-zero with
parameters
\begin{align}
w_{k+1|k+1}^{i,a^{i}} & =r_{k+1|k}^{i,\widetilde{a}^{i}}p^{D}\left(\beta_{k+1|k}^{i,\widetilde{a}^{i}}\left(1\right)+\beta_{k+1|k}^{i,\widetilde{a}^{i}}\left(2\right)\gamma\right)\nonumber \\
 & \quad\times\mathcal{N}(Z;k,\overline{z}^{i,a^{i}}(1),S_{i,a^{i}}(1))\label{eq:weight_died_upd}\\
\beta_{k+1|k+1}^{i,a^{i}}(l) & \propto\begin{cases}
\beta_{k+1|k}^{i,\widetilde{a}^{i}}\left(1\right) & l=1\\
\beta_{k+1|k}^{i,\widetilde{a}^{i}}\left(2\right)\gamma & l=2
\end{cases}\label{eq:beta_died_upd}\\
\overline{z}^{i,a^{i}}(l) & =H_{1,\tau}\overline{x}_{k+1|k}^{i,\widetilde{a}^{i}}(l)\label{eq:z_died_upd}\\
S_{i,a^{i}}(l) & =H_{1,\tau}P_{k+1|k}^{i,\widetilde{a}^{i}}(l)H_{1,\tau}^{T}+R_{1}\label{eq:S_died_upd}\\
\overline{x}_{k+1|k+1}^{i,a^{i}}(l) & =\overline{x}_{k+1|k}^{i,\widetilde{a}^{i}}(l)\nonumber \\
 & \quad+P_{k+1|k}^{i,\widetilde{a}^{i}}(l)H_{1,\tau}^{T}S_{i,a^{i}}^{-1}(l)\left(z_{1}-\overline{z}^{i,a^{i}}(l)\right)\label{eq:mean_died_upd}\\
P_{k+1|k+1}^{i,a^{i}}(l) & =P_{k+1|k}^{i,\widetilde{a}^{i}}(l)\nonumber \\
 & \quad-P_{k+1|k}^{i,\widetilde{a}^{i}}(l)H_{1,\tau}^{T}S_{i,a^{i}}^{-1}(l)H_{1,\tau}P_{k+1|k}^{i,\widetilde{a}^{i}}(l),\label{eq:cov_died_upd}
\end{align}
where $\tau=1$ for $l=1$ and $\tau=3$ for $l=2$.

If $Z^{j}=(k,z_{1:2})\in M_{(k+1)}^{3}$, then based on (\ref{eq:l_case1}),
it follows that $\beta_{k+1|k+1}^{i,a^{i}}(1)=0$ in (\ref{eq:Gaus_single_trj_density}),
and the only non-zero component in the mixture (\ref{eq:Gaus_single_trj_density})
corresponds to $l=2$ with parameters 
\begin{align}
w_{k+1|k+1}^{i,a^{i}} & =r_{k+1|k}^{i,\widetilde{a}^{i}}p_{D}\beta_{k+1|k}^{i,\widetilde{a}^{i}}(2)\widetilde{p}^{D}\mathcal{N}(Z;k,\overline{z}^{i,a^{i}},S_{i,a^{i}})\label{eq:weigth_alive_upd}\\
\beta_{k+1|k+1}^{i,a^{i}}(2) & =1\label{eq:beta_alive_upd}\\
\overline{z}^{i,a^{i}} & =H_{3,3}\overline{x}_{k+1|k}^{i,\widetilde{a}^{i}}(2)\label{eq:z_alive_upd}\\
S_{i,a^{i}} & =H_{3,3}P_{2,k+1|k}^{i,\widetilde{a}^{i}}H_{3,3}^{T}+R_{2}\label{eq:S_alive_upd}\\
\overline{x}_{k+1|k+1}^{i,a^{i}}(2) & =\overline{x}_{k+1|k}^{i,\widetilde{a}^{i}}(2)\nonumber \\
 & \quad+P_{k+1|k}^{i,\widetilde{a}^{i}}(2)H_{3,3}^{T}S_{i,a^{i}}^{-1}\left(z_{1:2}-\overline{z}^{i,a^{i}}\right)\label{eq:mean_alive_upd}\\
P_{k+1|k+1}^{i,a^{i}}(2) & =P_{k+1|k}^{i,\widetilde{a}^{i}}(2)\nonumber \\
 & \quad-P_{k+1|k}^{i,\widetilde{a}^{i}}(2)H_{3,3}^{T}S_{i,a^{i}}^{-1}H_{3,3}P_{k+1|k}^{i,\widetilde{a}^{i}}(2).\label{eq:cov_alive_upd}
\end{align}

If $Z^{j}=(k+1,z_{1})\in M_{(k+1)}^{2}$, then based on (\ref{eq:l_case3}),
it follows that $\beta_{k+1|k+1}^{i,a^{i}}(1)=0$ in (\ref{eq:Gaus_single_trj_density}),
and the only non-zero component in the mixture (\ref{eq:Gaus_single_trj_density})
corresponds to $l=2$ with parameters 
\begin{align}
w_{k+1|k+1}^{i,a^{i}} & =r_{k+1|k}^{i,\widetilde{a}^{i}}p^{D}\beta_{k+1|k}^{i,\widetilde{a}^{i}}(2)\gamma\mathcal{N}(Z;k+1,\overline{z}^{i,a^{i}},S_{i,a^{i}})\label{eq:weigth_new_upd}\\
\beta_{k+1|k+1}^{i,a^{i}}(2) & =1\label{eq:beta_new_upd}\\
\overline{z}^{i,a^{i}} & =H_{2,3}\overline{x}_{k+1|k}^{i,\widetilde{a}^{i}}(2)\label{eq:z_new_upd}\\
S_{i,a^{i}} & =H_{2,3}P_{k+1|k}^{i,\widetilde{a}^{i}}(2)H_{2,3}^{T}+R_{1}\label{eq:S_new_upd}\\
\overline{x}_{k+1|k+1}^{i,a^{i}}(2) & =\overline{x}_{k+1|k}^{i,\widetilde{a}^{i}}(2)\nonumber \\
 & \quad+P_{k+1|k}^{i,\widetilde{a}^{i}}(2)H_{2,3}^{T}S_{i,a^{i}}^{-1}\left(z_{1}-\overline{z}^{i,a^{i}}\right)\label{eq:mean_new_upd}\\
P_{k+1|k+1}^{i,a^{i}}(2) & =P_{k+1|k}^{i,\widetilde{a}^{i}}(2)\nonumber \\
 & \quad-P_{k+1|k}^{i,\widetilde{a}^{i}}(2)H_{2,3}^{T}S_{i,a^{i}}^{-1}H_{2,3}P_{k+1|k}^{i,\widetilde{a}^{i}}(2).\label{eq:cov_new_upd}
\end{align}
For a new Bernoulli component $f_{k+1|k+1}^{i,a^{i}}(\cdot)$, $i\in\{n_{k+1|k}+j\}$,
$j\in\{1,\dots,m_{k+1}\}$, initiated by the trajectory measurement
$Z^{j}=(t,z_{1:\nu})\in M_{(k+1)}^{\mu}$, $\mu\in\{1,2,3\}$, the
first hypothesis expresses a missed detection and has parameters $w_{k+1|k+1}^{i,1}=1$,
$r_{k+1|k+1}=0$. For the second hypothesis, we first calculate $\left\langle \lambda_{k+1|k},l\left(Z^{j}|\cdot\right)p^{D}\right\rangle $
in (\ref{eq:tracklets:w_gen_upd_new}) for each intensity component
in the two mixtures (\ref{eq:PPP_update-1}). The result of the inner
product $\left\langle \lambda_{k+1|k}^{B},l\left(Z^{j}|\cdot\right)p^{D}\right\rangle $
involving the $q_{b}$-th component of the first mixture in (\ref{eq:PPP_update-1})
is denoted as $v^{b,q_{b}}$, and the result of the $q_{p}$-th component
from the second mixture in (\ref{eq:PPP_update-1}) is denoted as
$v^{p,q_{p}}$. For a component of the first mixture, we have\footnote{Note that the weight of the Poisson components is not included in
(58) in \cite{GarciaFernandez2020a} due to a typographical error.
It is correctly included in (\ref{eq:tracklets:inner_prod_b}) and
(\ref{eq:tracklets:v_p}).}
\begin{align}
v^{b,q_{b}} & =w_{k+1|k}^{b,q_{b}}p^{D}\mathcal{N}(z_{1};H_{2,2}\overline{x}_{k+1|k}^{b,q_{b}},S_{b,q_{b}})\label{eq:tracklets:inner_prod_b}\\
S_{b,q_{b}} & =H_{2,2}P_{k+1|k}^{b,q_{b}}H_{2,2}^{T}+R_{1},\label{eq:tracklets:cov_b_inner_prod}
\end{align}
and, for a component of the second mixture, the inner product is
\begin{align}
 & v^{p,q_{p}}=\nonumber \\
 & \begin{cases}
w_{k+1|k}^{p,q_{p}}p^{D}\gamma\mathcal{N}(z_{1};H_{\widetilde{\mu},3}\overline{x}_{k+1|k}^{p,q_{p}},S_{p,q_{p},\widetilde{\mu}}) & Z^{j}\in M_{(k+1)}^{\widetilde{\mu}}\\
w_{k+1|k}^{p,q_{p}}p^{D}\widetilde{p}^{D}\mathcal{N}(z_{2};H_{3,3}\overline{x}_{k+1|k}^{p,q_{p}},S_{p,q_{p},3}), & Z^{j}\in M_{(k+1)}^{3},
\end{cases}\label{eq:tracklets:v_p}
\end{align}
where $\widetilde{\mu}\in\{1,2\}$ and
\begin{align}
S_{p,q_{p},\mu} & =\begin{cases}
H_{\mu,3}P_{k+1|k}^{b,q_{b}}H_{\mu,3}^{T}+R_{1} & \mu=\widetilde{\mu}\in\{1,2\}\\
H_{3,3}P_{k+1|k}^{b,q_{b}}H_{3,3}^{T}+R_{2} & \mu=3.
\end{cases}\label{eq:tracklets:S_partials}
\end{align}
Then, we compute $q_{b}^{*}=\max_{q_{b}}\left(v^{b,q_{b}}\right)$,
$q_{p}^{*}=\max_{q_{p}}\left(v^{p,q_{p}}\right)$ and set
\begin{align}
w_{k+1|k+1}^{i,2} & =\lambda^{C}\left(Z^{j}\right)+\sum_{q_{b}=1}^{n_{k+1|k}^{b}}v^{b,q_{b}}+\sum_{q_{p}=1}^{n_{k+1}^{p}}v^{p,q_{p}}\label{eq:W_newB_Gauss}\\
r_{k+1|k+1}^{i,2} & =\frac{\sum_{q_{b}=1}^{n_{k+1|k}^{b}}v^{b,q_{b}}+\sum_{q_{p}=1}^{n_{k+1}^{p}}v^{p,q_{p}}}{w_{k+1|k+1}^{i,2}}\label{eq:R_newB_Gauss}\\
p_{k+1|k+1}^{i,2}(X_{k+1}) & =\mathcal{N}\left(X_{k+1};t^{i,2},\overline{x}_{k+1|k+1}^{i,2},P_{k+1|k+1}^{i,2}\right).\label{eq:Gaussian_state_update_newB}
\end{align}
If \textbf{$v^{b,q_{b}^{*}}>v^{p,q_{p}^{*}}$,} $t^{i,2}=k+1$ and
we set
\begin{align}
\overline{x}_{k+1|k+1}^{i,2} & =\overline{x}_{k+1|k}^{b,q_{b}^{*}}+P_{k+1|k}^{b,q_{b}^{*}}H_{2,2}S_{b,q_{b}^{*}}^{-1}(z_{1}-H_{2,2}\overline{x}_{k+1|k}^{b,q_{b}^{*}})\label{eq:mean_newB}\\
P_{k+1|k+1}^{i,2} & =P_{k+1|k}^{b,q_{b}^{*}}-P_{k+1|k}^{b,q_{b}^{*}}H_{2,2}^{T}S_{b,q_{b}^{*}}^{-1}H_{2,2}P_{k+1|k}^{b,q_{b}^{*}},\label{eq:cov_newB}
\end{align}
based on (\ref{eq:l_case9}). If \textbf{$v^{b,q_{b}^{*}}<v^{p,q_{p}^{*}}$},
the measurement model takes the form of (\ref{eq:l_case1})-(\ref{eq:l_case3}),
$t^{i,2}=k$ and $\overline{x}_{k+1|k+1}^{i,2}$, $P_{k+1|k+1}^{i,2}$
are given by substituting $\overline{x}_{k+1}^{p,q_{p}^{*}}$ and
$P_{k+1}^{p,q_{p}^{*}}$ into (\ref{eq:mean_died_upd})-(\ref{eq:cov_died_upd}),
(\ref{eq:mean_alive_upd})-(\ref{eq:cov_alive_upd}) and (\ref{eq:mean_new_upd})-(\ref{eq:cov_new_upd})
for $Z^{j}\in M_{(k+1)}^{1}$, $Z^{j}\in M_{(k+1)}^{3}$ and $Z^{j}\in M_{(k+1)}^{2}$,
respectively.
\end{lem}
The proof of Lemma \ref{lem:Gaus_TMPMBM_upd} is given in Appendix
\ref{sec:Proof-of-Lemma5}. Note that if the measurement is detected
only at time step $k$, the component correspondent to $l=1$ in (\ref{eq:Gaus_single_trj_density}),
defined by (\ref{eq:beta_died_upd})-(\ref{eq:cov_died_upd}), will
be discarded after the marginalisation step, and that $\beta_{k+1|k}^{i,\widetilde{a}^{i}}\left(1\right)+\beta_{k+1|k}^{i,\widetilde{a}^{i}}\left(2\right)=1$.
Furthermore, from (\ref{eq:single_state_density_gen_upd_new-1}),
the single trajectory density of the new Bernoulli components is a
Gaussian mixture of two-step trajectories with start times $t^{i,2}\in\{k,k+1\}$.
To improve the filter's computational efficiency, we use a Gaussian
approximation by defining (\ref{eq:Gaussian_state_update_newB}) based
on the Gaussian component with the highest weight in the mixture.

\subsection{Marginalisation\label{subsec:Marginalisation_GM}}

\begin{algorithm}
\textbf{Input: }Parameters of the TM-PMBM posterior at the previous time step, see Sec. \ref{subsec:Prediction_GM}, and set of trajectory measurements $\mathbf{Z}_{k+1}$ at current time window.\\
\textbf{Output: }Parameters of the TM-PMBM posterior at the current time window.
\begin{algorithmic}
\State - Perform prediction, see Lemma \ref{lem:Gaus_TMPMBM_pred}.
\State - Update the PPP intensity (misdetection) (\ref{eq:PPP_update-1})-(\ref{eq:weight_PPP_upd}). 
\For{$Z\in\mathbf{Z}_{k+1}$} \Comment{Targets detected for the first time}
\State - Perform gating of $Z$ w.r.t. Gaussian components of 
\Statex \hspace{\algorithmicindent} \, Poisson prior (\ref{eq:PPP_GM}).
\If{$Z$ within the gate of at least one PPP component}
\State - Create a new Bernoulli component (\ref{eq:W_newB_Gauss})-(\ref{eq:cov_newB}) 
\Statex \hspace{\algorithmicindent}\hspace{\algorithmicindent}\, computing $v^{b,q_{b}^{*}}$ and $v^{p,q_{p}^{*}}$.
\EndIf
\EndFor
\For{$i=1$ to $n_{k+1|k}$} \Comment{Go through all possible targets}
\For{$j_i=1$ to $h^i$}\\ 
\Comment{$h^i$ is the number of single-target hypotheses for target $i$}
\State - Create new misdetection hypothesis (\ref{eq:W_mis_update})-(\ref{eq:cov_gaus_missD}).
\State - Perform gating on $Z$ and form detection hypotheses \Statex \hspace{\algorithmicindent}\hspace{\algorithmicindent}\, by trajectory type (\ref{eq:weight_died_upd})-(\ref{eq:cov_new_upd}).
\EndFor
\EndFor
\State - Form the updated global hypotheses, see \cite[Sec.~V-C3]{Garcia-Fernandez18}.
\State - Marginalise PPP intensity (\ref{eq:PPP_GM_marg}) and Bernoulli components\Statex \, (\ref{eq:prob_ext_GM_marg})-(\ref{eq:mean_cov_gaus_GM_marg}).
\State - Estimate target states by selecting the global hypothesis \Statex \, with the highest weight, see \cite[Sec.~VI-A]{Garcia-Fernandez18}.
\State - Prune low-weigth Poisson components, low-weight global \Statex \, hypotheses and Bernulli components with low probability \Statex \, of existence.
\end{algorithmic}\caption{Pseudocode of the TM-PMBM filter.\label{alg:TM_PMBM_pseoudocode}}
\end{algorithm}
Following Lemma \ref{lem:TM-PMBM marginalisation}, we compute the
marginalisation step in the Gaussian implementation.
\begin{lem}
[Gaussian TM-PMBM marginalisation]\label{lem:Gaus_TMPMBM_marg}Assume
the updated density at time step $k$ and $k+1$ is a PMBM on the
set of two-step trajectories of the form (\ref{eq:PMBM_on_trj}),
with $\lambda_{k+1|k+1}(\cdot)$ of the form (\ref{eq:PPP_GM}) and
Bernoulli components $f_{k+1|k+1}^{i,a^{i}}(\cdot)$, $i\in\{1,\dots,n_{k+1|k+1}\}$,
$a^{i}\in\{1,\dots,h_{k+1|k+1}^{i}\}$, with single trajectory density
$p_{k+1|k+1}^{i,a^{i}}(\cdot)$ of the form (\ref{eq:Gaus_single_trj_density}).
The updated density over the set of target states is a PMBM of the
form (\ref{eq:PMBM_on_states}), with PPP intensity
\begin{align}
\lambda_{k+1|k+1}^{M}(x_{k+1}) & =(1-p^{D})\sum_{q_{b}=1}^{n_{k+1}^{b}}w_{k+1}^{b,q_{b}}\mathcal{N}(x_{k+1};\overline{x}_{k+1}^{b,q_{b}},P_{k+1}^{b,q_{b}})\nonumber \\
 & +\sum_{q_{p}=1}^{n_{k+1|k}^{p}}w_{k+1|k+1}^{p,q_{p}}\mathcal{N}(x_{k+1};\overline{x}_{k+1|k}^{p,q_{p},M},P_{k+1|k}^{p,q_{p},M}),\label{eq:PPP_GM_marg}
\end{align}
where $\overline{x}_{k+1|k}^{p,q_{p},M}=\overline{x}_{k+1|k}^{p,q_{p},2}\in\mathbb{R}^{n_{x}}$
and $P_{k+1|k}^{p,q_{p},M}=P_{k+1|k}^{p,q_{p},2,2}\in\mathbb{R}^{n_{x}\times n_{x}}$
are the mean vector and covariance matrix at time step $k+1$, see
(\ref{eq:mean_upd_componets}) and (\ref{eq:covar_upd_components}).
Based on Lemma \ref{lem:TM-PMBM marginalisation}, the Bernoulli components
of the posterior density after the marginalisation $f_{k+1|k+1}^{i,a^{i},M}(\cdot)$,
$i\in\{1,\dots,n_{k+1|k+1}\}$, $a^{i}\in\{1,\dots,h_{k+1|k+1}^{i}\}$
have probability of existence 
\begin{equation}
r_{k+1|k+1}^{i,a^{i},M}=r_{k+1|k+1}^{i,a^{i}}\beta_{k+1|k+1}^{i,a^{i}}(2),\label{eq:prob_ext_GM_marg}
\end{equation}
and single-target density 
\begin{equation}
p_{k+1|k+1}^{i,a^{i},M}(x_{k+1})=\mathcal{N}(x_{k+1};\overline{x}_{k+1|k+1}^{i,a^{i},M},P_{k+1|k+1}^{i,a^{i},M}),\label{eq:single-trarget_density_marg}
\end{equation}
where
\begin{align}
\overline{x}_{k+1|k+1}^{i,a^{i},M} & =\overline{x}_{k+1|k+1}^{i,a^{i},2}(2),\quad & P_{k+1|k+1}^{i,a^{i},M} & =P_{k+1|k+1}^{i,a^{i},2,2}(2).\label{eq:mean_cov_gaus_GM_marg}
\end{align}
\end{lem}

\subsection{TM-PMB filter\label{sec:tracklets:TM-PMB-filter}}

\begin{figure}
\centering{}\includegraphics[scale=0.37]{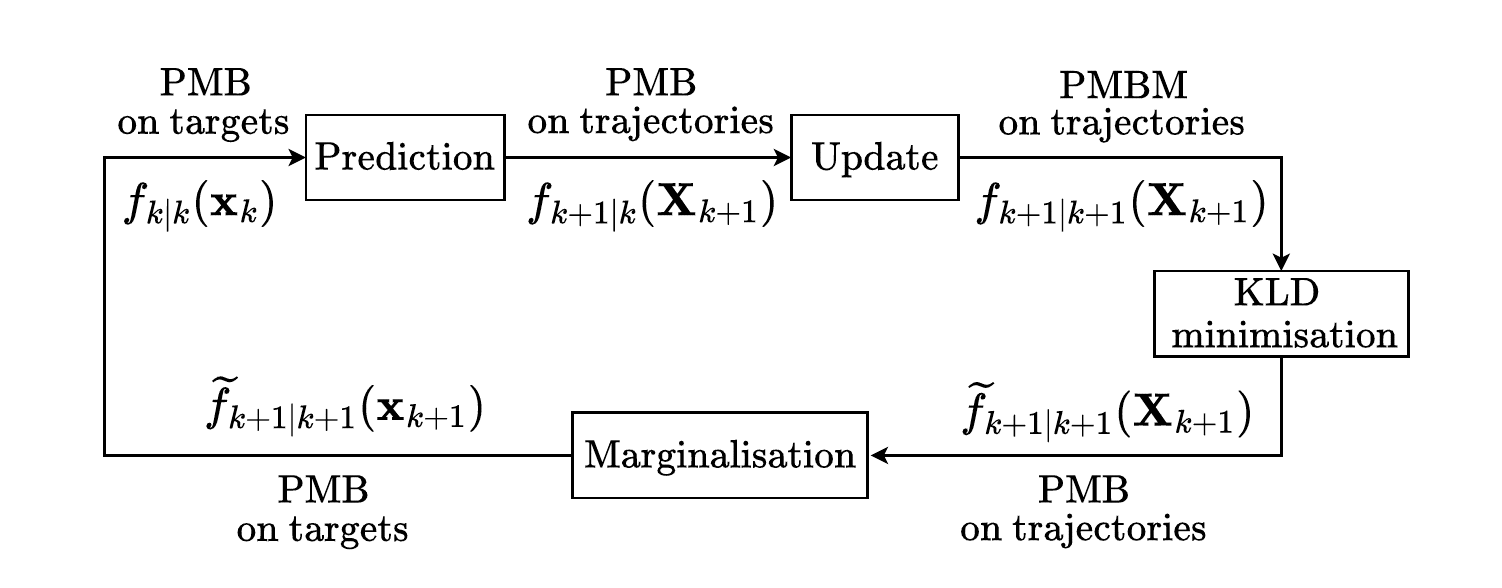}\caption{Diagram of the PMB filter based on trajectory measurements, the TM-PMB
filter. The PMB density on the two-step trajectories is computed via
KLD minimisation (with auxiliary variables \cite{GarciaFernandez2020a})
of the PMBM posterior obtained from the update step performed with
trajectory measurements. The order of the KLD minimisation and the
marginalisation step can be swapped without affecting the result.\label{fig:tracklets:PMB_schematic}}
\end{figure}

In this section, we discuss the (track-oriented) TM-PMB filter, which
operates similarly to the PMB filter \cite{Williams15a}. It can be
derived through KLD minimisation by introducing auxiliary variables
in the update step, as in the derivation of the trajectory PMB filter
\cite{GarciaFernandez2020a}. Fig. \ref{fig:tracklets:PMB_schematic}
provides a diagram illustrating the operation of the TM-PMB filter.

The resulting equations for the KLD minimisation step are
\begin{align}
\widetilde{\lambda}_{k|k}\left(X\right) & =\lambda_{k|k}\left(X\right)\\
r_{k|k}^{i} & =\sum_{a^{i}=1}^{h_{k|k}^{i}}\overline{w}_{k|k}^{i,a^{i}}r_{k|k}^{i,a^{i}}\label{eq:existence_Prop2}\\
p_{k|k}^{i}\left(X\right) & =\frac{\sum_{a^{i}=1}^{h_{k|k}^{i}}\overline{w}_{k|k}^{i,a^{i}}r_{k|k}^{i,a^{i}}p_{k|k}^{i,a^{i}}\left(X\right)}{\sum_{a^{i}=1}^{h_{k|k}^{i}}\overline{w}_{k|k}^{i,a^{i}}r_{k|k}^{i,a^{i}}},\label{eq:p_i_Prop2}
\end{align}
where
\begin{align}
\overline{w}_{k|k}^{i,a^{i}} & =\sum_{b\in\mathcal{A}_{k|k}:b^{i}=a^{i}}w_{k|k}^{b}.\label{eq:weight_simplified_Prop2}
\end{align}
After completing the minimisation step, the marginalisation step described
in Section \ref{subsec:Marginalisation} is performed to obtain the
PMB posterior density on the set of target states. 

\section{Simulations\label{sec:tracklets:Simulations}}

In this section, we assess the performance of the TM-PMBM filter\footnote{The MATLAB implementation will be made publicly available at \mbox{https://github.com/mrcfon}.}
and compare it with the standard PMBM filter based on sets of targets
\cite{Garcia-Fernandez18}, as well as the corresponding PMB approximations
\cite{Williams15a}. The PMBM and PMB filters on sets of targets (which
we simply refer to as PMBM and PMB filters) process only a set of
(standard) measurements at each time step. For that reason, they discard
the part of the trajectory measurements that corresponds to the time
step at the beginning of the time window and proceed with an adjusted
probability of detection, $\overline{p}^{D}$ and clutter intensity
$\lambda_{2}^{C}\left(\cdot\right)$. All units in this section are
presented in the International System of Units (SI) for clarity in
notation and they are omitted for brevity.

We have also tested PMBM/PMB filters that process point measurements
obtained from decomposing the trajectory measurements (without removing
any states). Nevertheless, this option implies that a target may generate
at most two point detections at each time step and standard point-target
PMBM/PMB filters do not work well in this setting (mainly due to an
increase in false targets). Therefore, for clarity, the results from
these PMBM/PMB filters are omitted. 

The filter implementations use a threshold for pruning the Poisson
components $\Gamma_{p}=10^{-5}$, a threshold for pruning global hypotheses
$\Gamma_{mbm}=10^{-4}$, and a threshold for pruning Bernoulli components
$\Gamma_{b}=10^{-5}$. The maximum number of global hypotheses is
set to $N_{h}=200$, and ellipsoidal gating is applied independently
to the both ends of the trajectory measurements with a threshold of
$\Gamma_{g}=9$. In the trajectory-based filters, a trajectory measurement
is considered for the update if at least one of its locations falls
within the respective gate. The estimation is carried out by selecting
the global hypothesis with the highest weight and and reporting Bernoulli
components that have an existence probability greater than 0.1 \cite[Sec. VI.A]{Garcia-Fernandez18}.
These parameters have been empirically determined to achieve good
performance, representing a reasonable trade-off between computational
burden and accuracy. All filters were implemented using Murty's algorithm
\cite{Murty1968}.

The evaluation is conducted in terms of root mean square (RMS) generalised
optimal subpattern assignment metric (GOSPA) error ($\alpha=2$, $c=10$,
$p=2$) \cite{Rahmathullah17}, which allows for the decomposition
of the total error into localisation error, missed target error and
false target error. We examine two scenarios characterised by different
parameters and structures, depicted in Fig. \ref{fig:tracklets:Scenario1_GT}
and \ref{fig:tracklets:Scenario2_GT}. For each scenario, we conduct
multiple simulations with different probabilities $\widetilde{p}^{D}$
of receiving a full trajectory measurement given the detection of
a full measurement, and varying values of mean number of clutter trajectory
measurements per scan $\overline{\lambda}^{C}$. These simulations
are evaluated across four different lengths of the time window $N_{w}$,
specifically $N_{w}=\left\{ 2,5,7,10\right\} $ time steps.

The simulations were conducted on a laptop featuring an Intel (R)
Core(TM) i7-8850H clocked at 2.60 GHz and 16 GB RAM. The implementation
utilised MATLAB for all components except Murty\textquoteright s algorithm,
which was implemented in C++\footnote{We used the Murty's algorithm implementation in the tracker component
library \cite{Crouse2017}.}. The reported results are averaged over 100 Monte Carlo (MC) runs.

\subsection{Models}

In the simulations, the motion of the targets follows a nearly constant
velocity model \cite{BarShalom2001}, and the target states are sampled
at a sampling interval of $T=0.2$. The single target state is defined
within a two-dimensional Cartesian coordinate system as $[p_{x,k},v_{x,k},p_{y,k},v_{y,k}]^{T}$,
where the first two components denote the position and velocity along
the $x$-axis, and the last two denote those along the $y$-axis.
In the filters, the parameters for the linear and Gaussian motion
and measurement models are as follows:
\begin{align}
F & =I_{2}\otimes\begin{pmatrix}1 & \Delta\\
0 & 1
\end{pmatrix} & Q & =qI_{2}\otimes\begin{pmatrix}\Delta^{3}/3 & \Delta^{2}/2\\
\Delta^{2}/2 & \Delta
\end{pmatrix}\label{eq:tracklets:trans_model_matrices}\\
H & =I_{2}\otimes\begin{pmatrix}1 & 0\end{pmatrix} & R & =\sigma^{2}I_{2}\,,\label{eq:tracklets:meas_model_matrices}
\end{align}
where $\otimes$ is the Kronecker product, $q=0.01$, $\sigma^{2}=0.1$,
and $\Delta$ is the time interval, defined as a multiple of the sampling
interval, i.e., $\Delta=TN_{w}$. In the following sections, we refer
to the scaling factor $N_{w}$ as the length of the time window to
indicate the results obtained by running the filters with different
time intervals $\Delta$.

We set the probability of detection to $p^{D}=0.9$ for the filters
based on trajectory measurements in all the simulations. For the filters
based on target states (PMBM/PMB), which only consider measurements
at the end of the time window, the equivalent probability of detection
is defined as $\overline{p}^{D}=p^{D}(1-\gamma)$. This represents
the probability of detection multiplied by the probability of cases
(\ref{eq:l_case1}) and (\ref{eq:l_case3}). Moreover, we set the
probability of survival to the next time step to $p^{S,T}=0.99$ and
the probability of survival to the next time window is $p^{S}=\left(p_{S}^{S,T}\right)^{N_{w}}$.

\subsubsection{Trajectory measurement clutter model\label{subsec:tracklets:Clutter-model}}

Clutter is a PPP on the trajectory measurement space $M_{(k+1)}$,
with clutter intensity is\begin{subnumcases}
{\lambda^{C}(Z_{k+1})=\label{eq:clutter_intensity}}
\overline{\lambda}_{P}^{C}f_{C}^{P,1}\left(Z_{k+1}\right) & $Z_{k+1}\in M_{(k+1)}^{1}$\\
\overline{\lambda}_{P}^{C}f_{C}^{P,2}\left(Z_{k+1}\right) & $Z_{k+1}\in M_{(k+1)}^{2}$\\
\overline{\lambda}_{F}^{C}f_{C}^{F}\left(Z_{k+1}\right) & \text{\ensuremath{Z_{k+1}\in M_{(k+1)}^{3},}}
\end{subnumcases}where $\overline{\lambda}_{F}^{C}$ is the clutter rate for a trajectory
measurement in $M_{(k+1)}^{3}$, and $\overline{\lambda}_{P}^{C}$
is the clutter rate for a trajectory measurement in $M_{(k+1)}^{1}$
or $M_{(k+1)}^{2}$. The spatial single measurement densities for
full and partial trajectory measurements are

\begin{subnumcases}
{f_{C}^{P,1}(k,z_{1}) = f_{C}^{P,2}(k+1,z_{1})=\label{eq:spatial_pdf_partial}}
\frac{1}{\left|V\right|} & $z_{1}\in V$\\
0 & otherwise,
\end{subnumcases}\begin{subnumcases}
{f_{C}^{F}(t,z_{1:2})=\label{eq:spatial_pdf_full}}
1/\left|V\right|^{2} & $z_{1:2}\in V\times V$\\
0 & otherwise,
\end{subnumcases}where $V$ is the area of the field of view of the sensor, and $|V|$
represents the volume/size of $V$. The spatial density (\ref{eq:spatial_pdf_partial})
is uniform in $V$, and the spatial pdf (\ref{eq:spatial_pdf_full})
is uniform in $V\times V$.

The clutter rate is the integral on the single-trajectory measurement
space \cite{GarciaFernandez2019a}
\begin{align}
\overline{\lambda}^{C}= & \int\lambda^{C}(Z_{k+1})\,dZ_{k+1}\label{eq:clutter_init_tot}\\
= & \int\lambda^{C}(k,z_{1:2})\,dz_{1:2}+\int\lambda^{C}(k,z_{1})\,dz_{1}\nonumber \\
 & \quad+\int\lambda^{C}(k+1,z_{1})\,dz_{1}\label{eq:clutter_decomp}\\
= & \overline{\lambda}_{F}^{C}+2\overline{\lambda}_{P}^{C}.\label{eq:clutter_rate_sum}
\end{align}
In the simulations, the clutter rates of full and partial clutter
trajectory measurements are equal, i.e, $\overline{\lambda}_{F}^{C}=\overline{\lambda}_{P}^{C}$.

The standard PMBM/PMB filters are defined on sets of targets. Therefore
the clutter model proposed in (\ref{eq:spatial_pdf_full}-\ref{eq:clutter_rate_sum})
cannot be used with these filters. We define the clutter model for
the PMBM/PMB filters by marginalising the PPP of the trajectory clutter
model at time step $k$. This results in \cite{Granstroem2019}
\begin{equation}
\lambda_{2}^{C}\left(z\right)=\left(\overline{\lambda}_{F}^{C}+\overline{\lambda}_{P}^{C}\right)f_{C}^{P,2}\left(k+1,z\right).\label{eq:tracklets:clutter_model_4_std}
\end{equation}

\subsection{Scenario 1\label{subsec:tracklets:Scenario-1}}

\begin{figure}
\centering{}%
\subfloat[Ground truth for Scenario 1, in which each target is represented by
a distinct colour.\label{fig:tracklets:Scenario1_GT}]%
{\centering{}\includegraphics[scale=0.2]{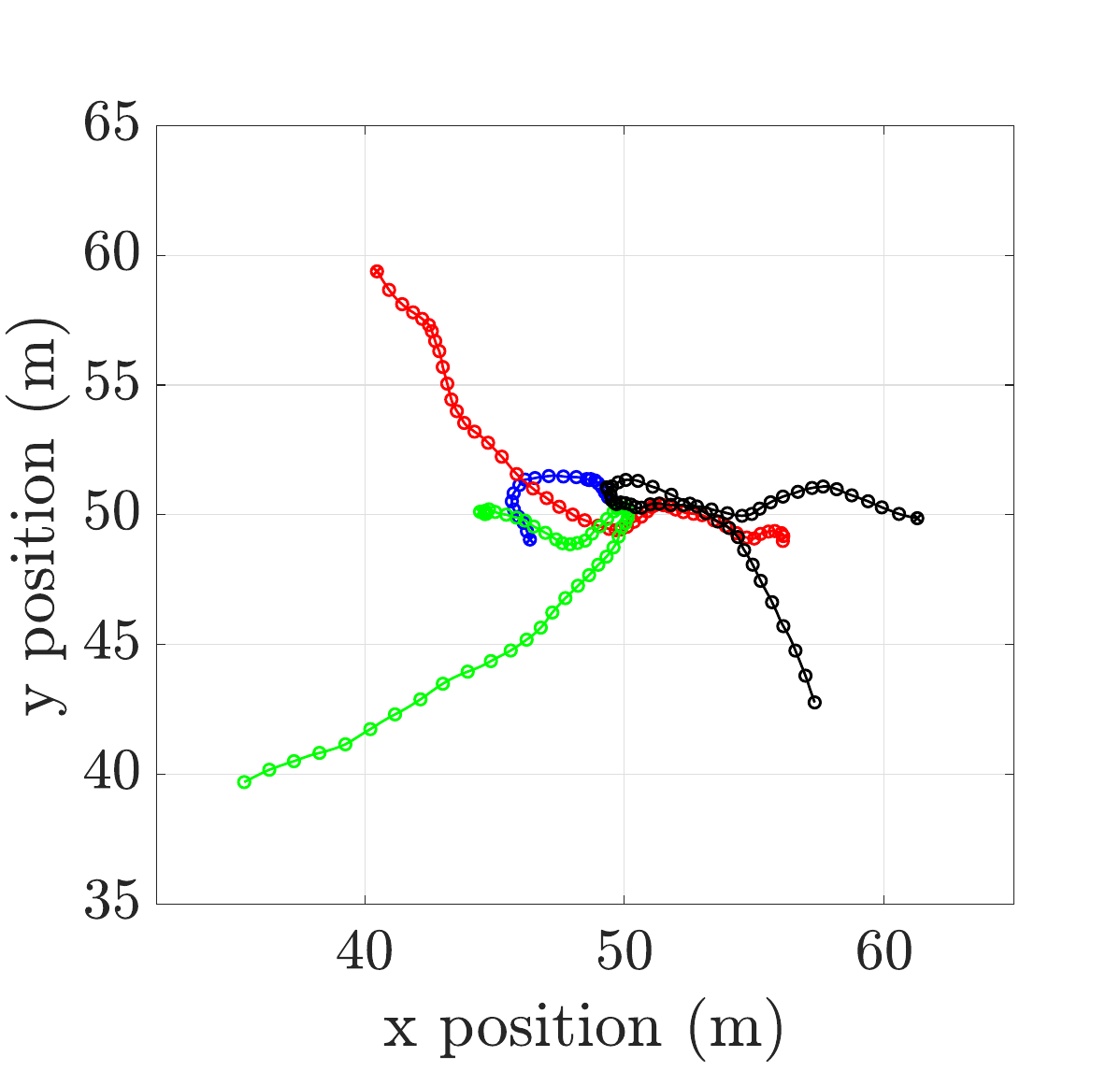}}%
\subfloat[Ground truth for Scenario 2, in which the colours represent the temporal
evolution of the target positions.\label{fig:tracklets:Scenario2_GT}]%
{\centering{}\includegraphics[scale=0.19]{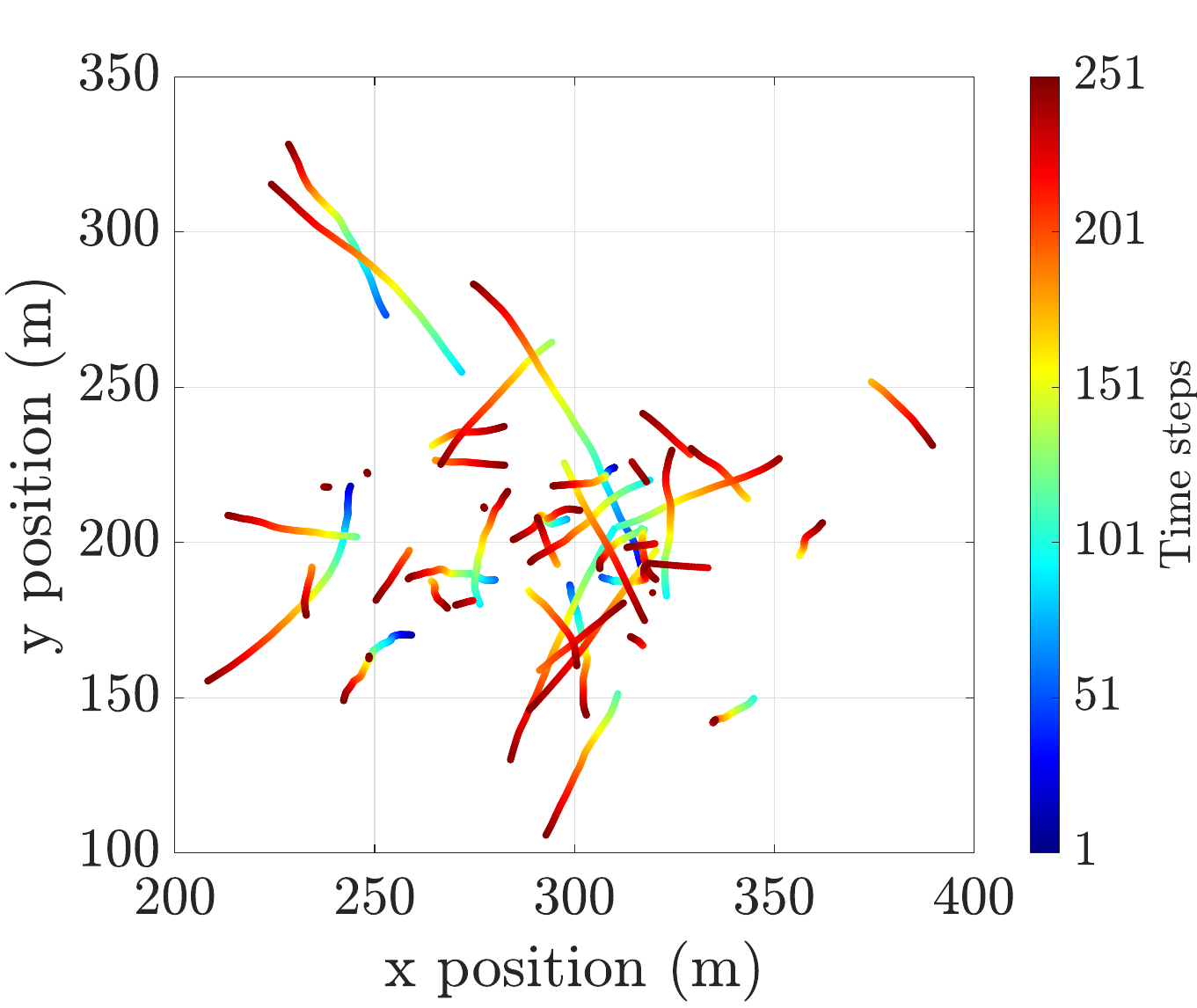}}\caption{Scenarios used to assess the performance of the TM-PMBM filter and
to compare it with the standard PMBM filter. In Fig. \ref{fig:tracklets:Scenario1_GT},
target positions at $k=1$ are marked with a cross, and subsequent
positions are indicated every five time steps with circles.\label{fig:Scenarios}}
\end{figure}
\begin{figure}
\centering{}\includegraphics[scale=0.33]{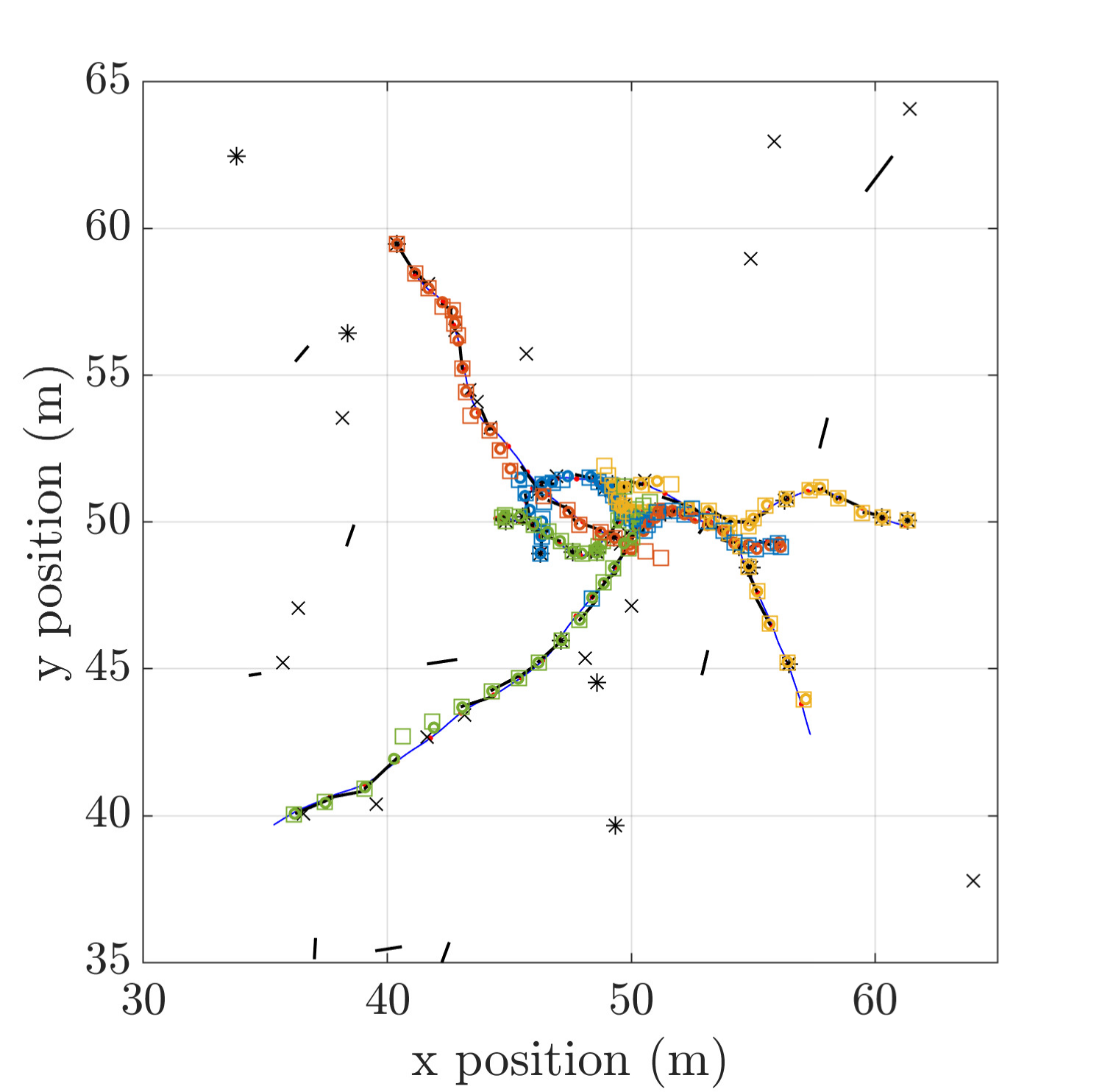}\caption{Comparison of the tracking performance of TM-PMBM (depicted with circle
markers) and standard PMBM (depicted with square markers) filters
based on Scenario 1, with $N_{w}=7$, $\widetilde{p}^{D}=0.7$ and
$\overline{\lambda}^{C}=10$. The ground-truth trajectories are shown
in blue, while the red dots denote the ground-truth locations at the
beginning and end of each time window. Measurements obtained across
all time steps are shown in black: $Z\in M_{(k+1)}^{1}$ with crosses,
$Z\in M_{(k+1)}^{2}$ with stars, and $Z=(k,z_{1:2})\in M_{(k+1)}^{3}$
with bars connecting $z_{1}$ and $z_{2}$. At many time steps, the
outputs of both filters are visually similar though at other time
steps, the TM-PMBM clearly outperforms the standard PMBM.\label{fig:Comparison-single_run}}
\end{figure}
Scenario 1 extends the base scenario outlined in \cite{Garcia-Fernandez18}.
It involves four targets, all originating at time step $k=1$ and
remaining alive throughout a simulation of 250 time steps, except
for one that dies at time step 125 (depicted in blue in Fig. \ref{fig:tracklets:Scenario1_GT}).
This scenario is recognised as challenging due to the convergence
of all targets around time step 125, when the blue one dies.

The targets are modelled as being born according to a PPP of intensity
$3$ at the first time step, and $0.005$ at the next time steps.
The intensity at each time step is Gaussian with mean $[50,0,50,0]^{T}$,
and covariance $\diag([50,1,50,1]^{2})$, which covers the considered
area of $[0,100]\times[0,100]$. In each time window, we approximate
the Gaussian mixture of new PPP components through Gaussian mixture
reduction, resulting in a single component that represents all possible
targets born within a time window.

\begin{figure}
\begin{centering}
\subfloat[\label{fig:tracklets:Williams_pd09_tot}]%
{\centering{}\includegraphics[scale=0.18]{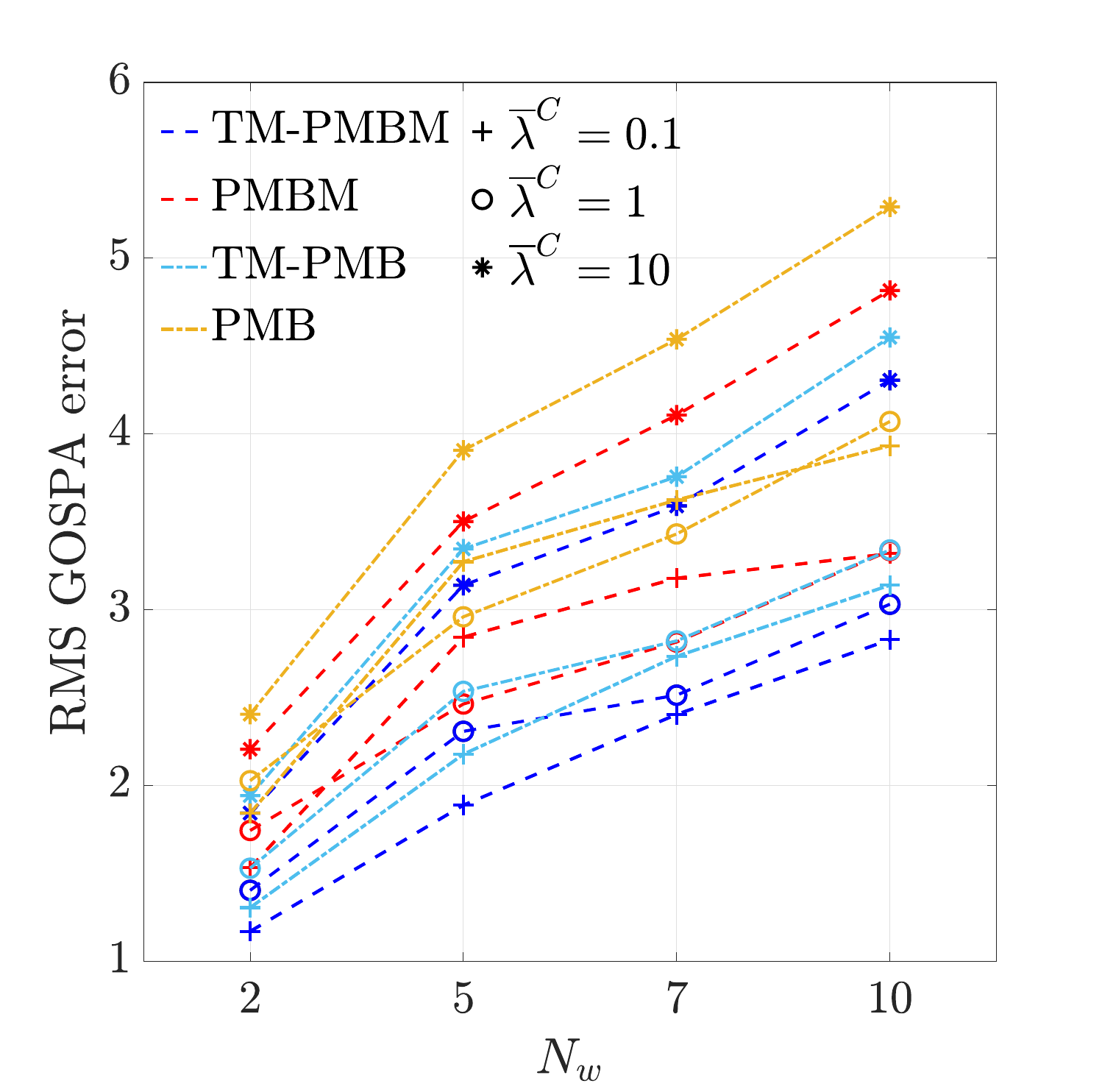}}%
\subfloat[\label{fig:tracklets:Williams_pd09_loc}]%
{\centering{}\includegraphics[scale=0.18]{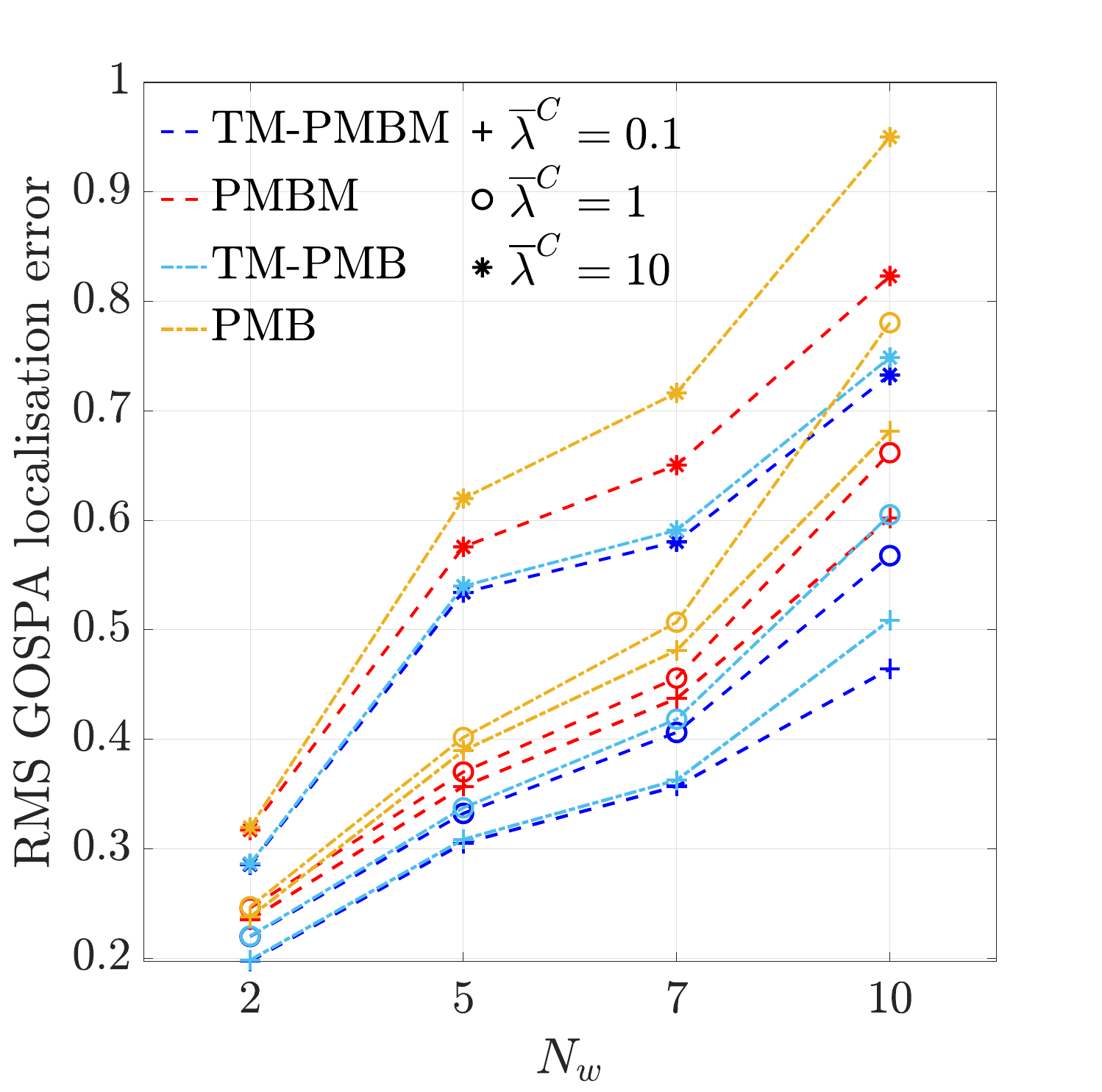}}\\
\subfloat[\label{fig:tracklets:Williams_pd09_false}]%
{\centering{}\includegraphics[scale=0.18]{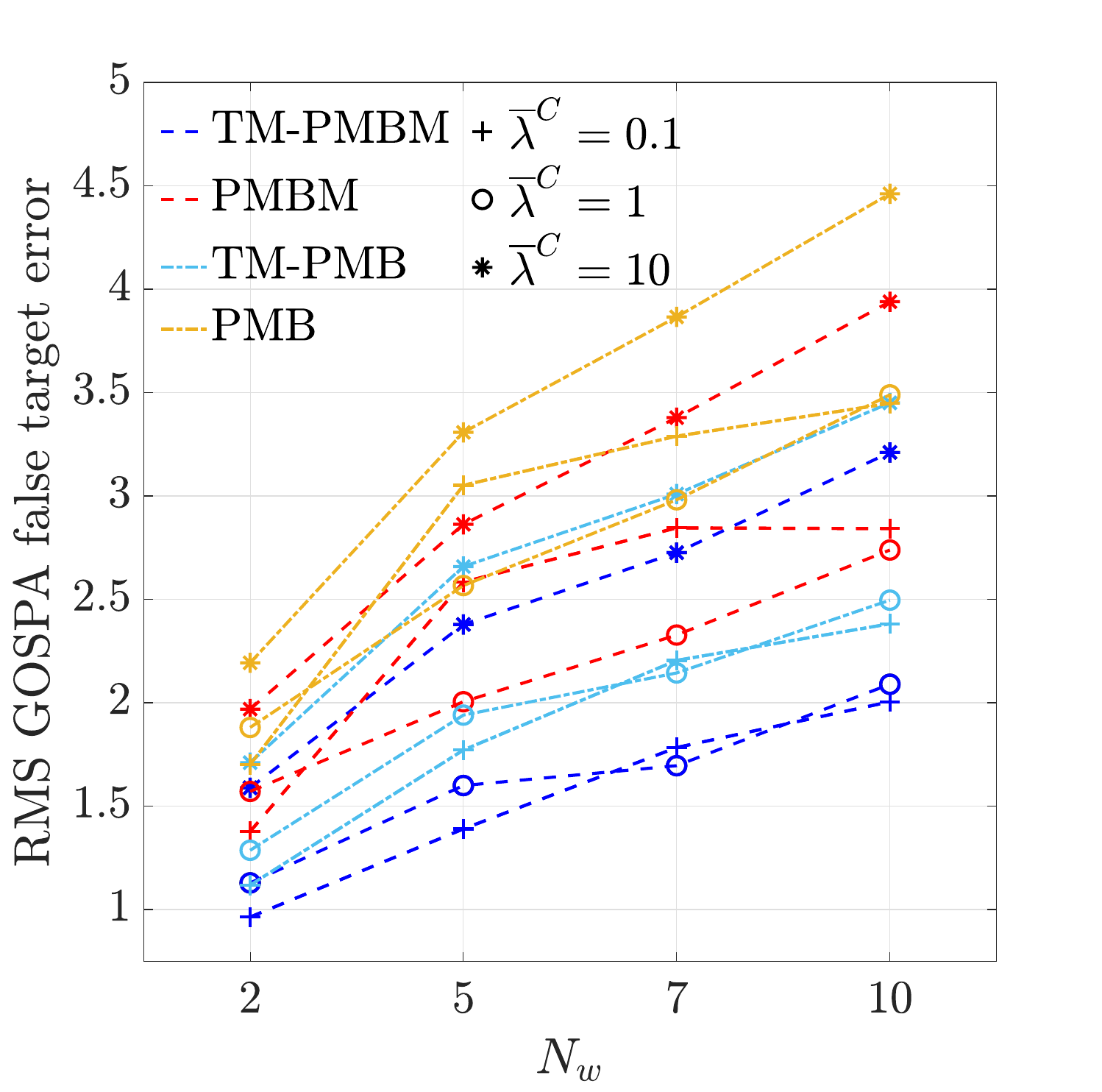}}%
\subfloat[\label{fig:tracklets:Williams_pd09_mis}]%
{\centering{}\includegraphics[scale=0.18]{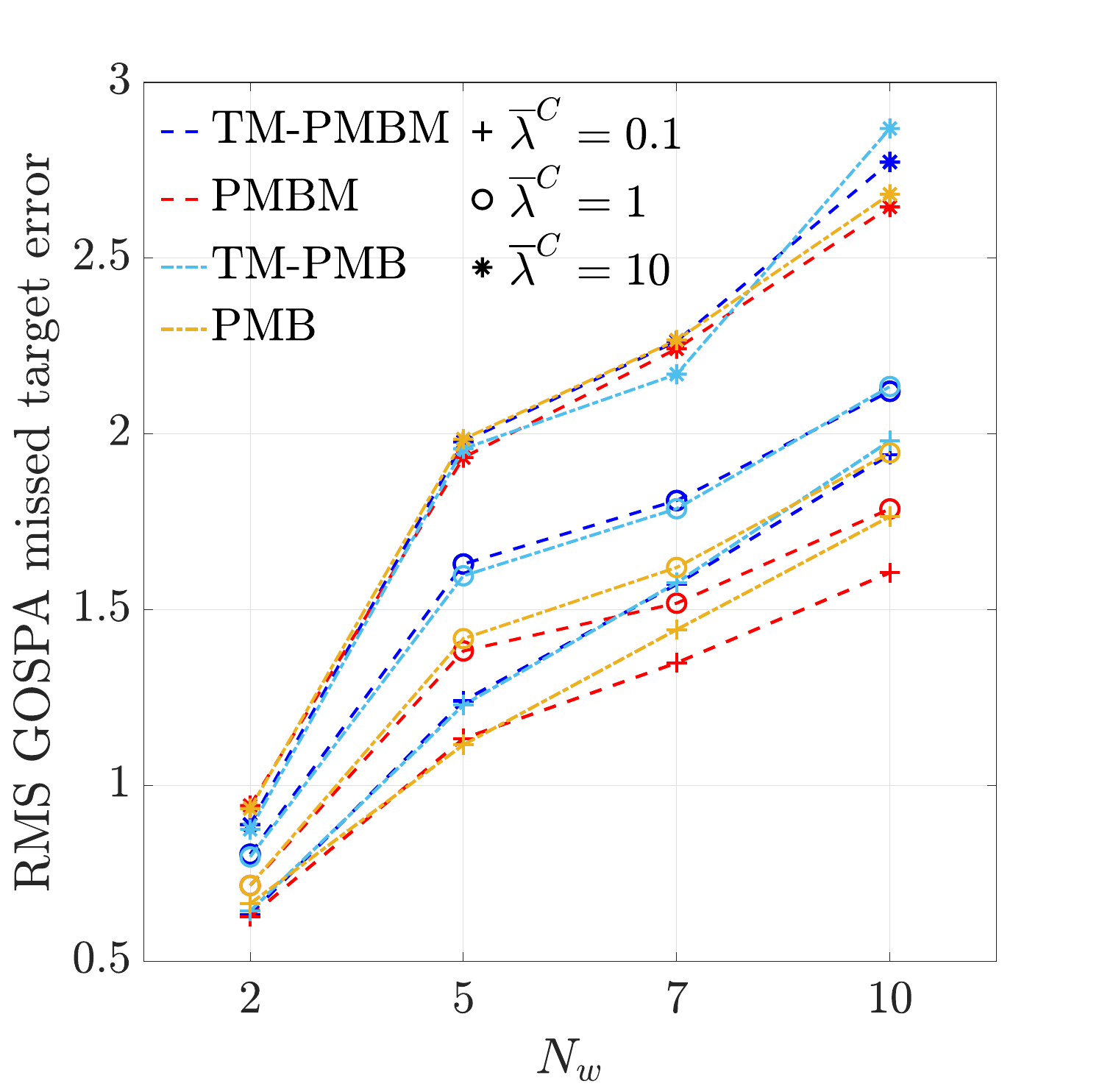}}
\par\end{centering}
\centering{}\caption{\label{fig:tracklets:Results_Williams_pd09}GOSPA metric results for
Scenario 1 with $\widetilde{p}^{D}=0.9$ and clutter rates $\overline{\lambda}^{C}=\{0.1,1,10\}$,
averaged over 100 MC runs. Each filter is represented by a distinct
colour, and simulations sharing the same clutter rate are identified
using the same marker. Results for the TM-PMBM and PMBM filters are
connected by dashed lines, while those for the TM-PMB and PMB filters
are connected by dash-dotted lines.}
\end{figure}

\begin{figure}
\begin{centering}
\subfloat[\label{fig:tracklets:Williams_pd07_tot}]%
{\centering{}\includegraphics[scale=0.18]{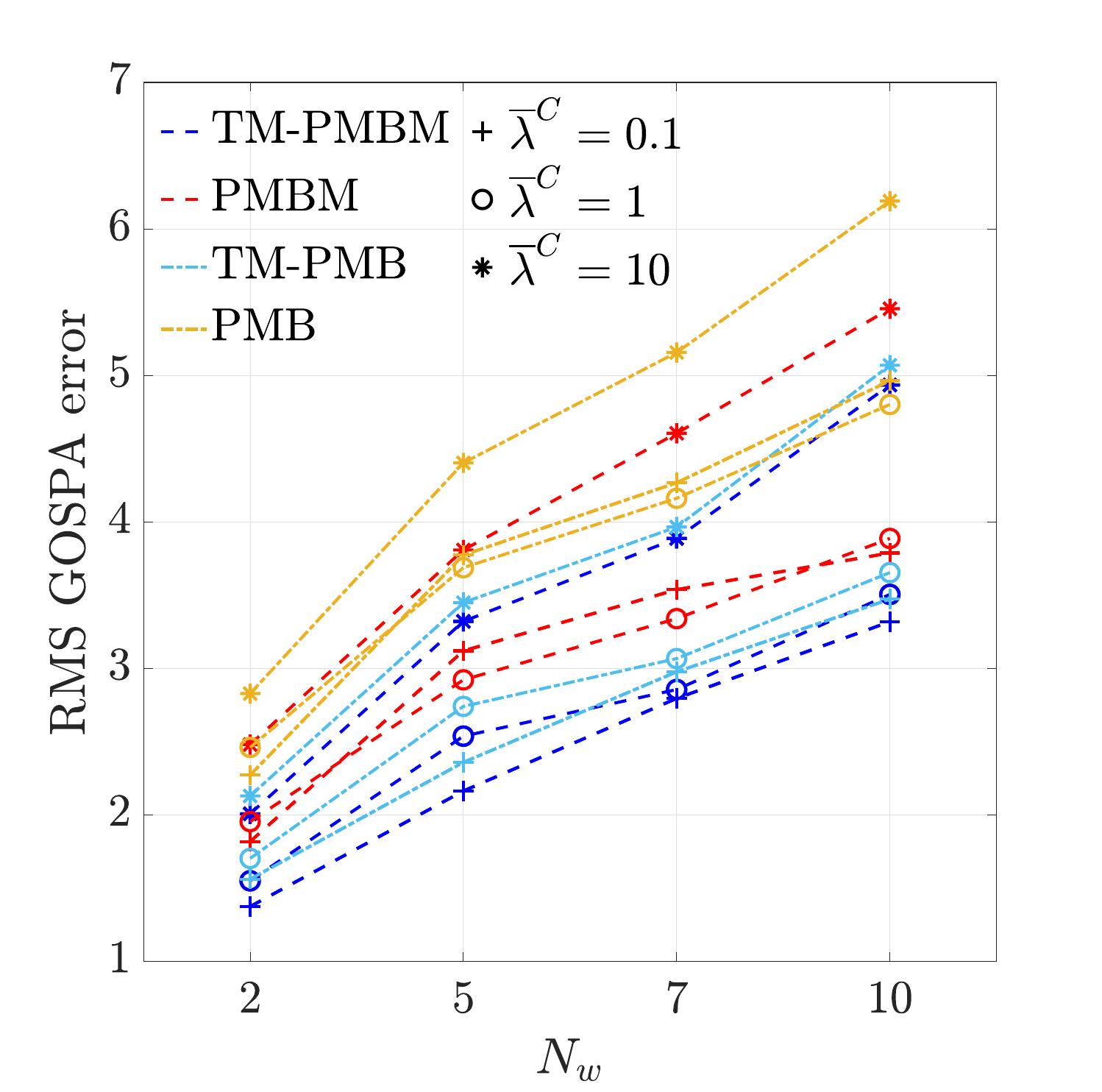}}%
\subfloat[\label{fig:tracklets:Willaims_pd07_loc}]%
{\centering{}\includegraphics[scale=0.18]{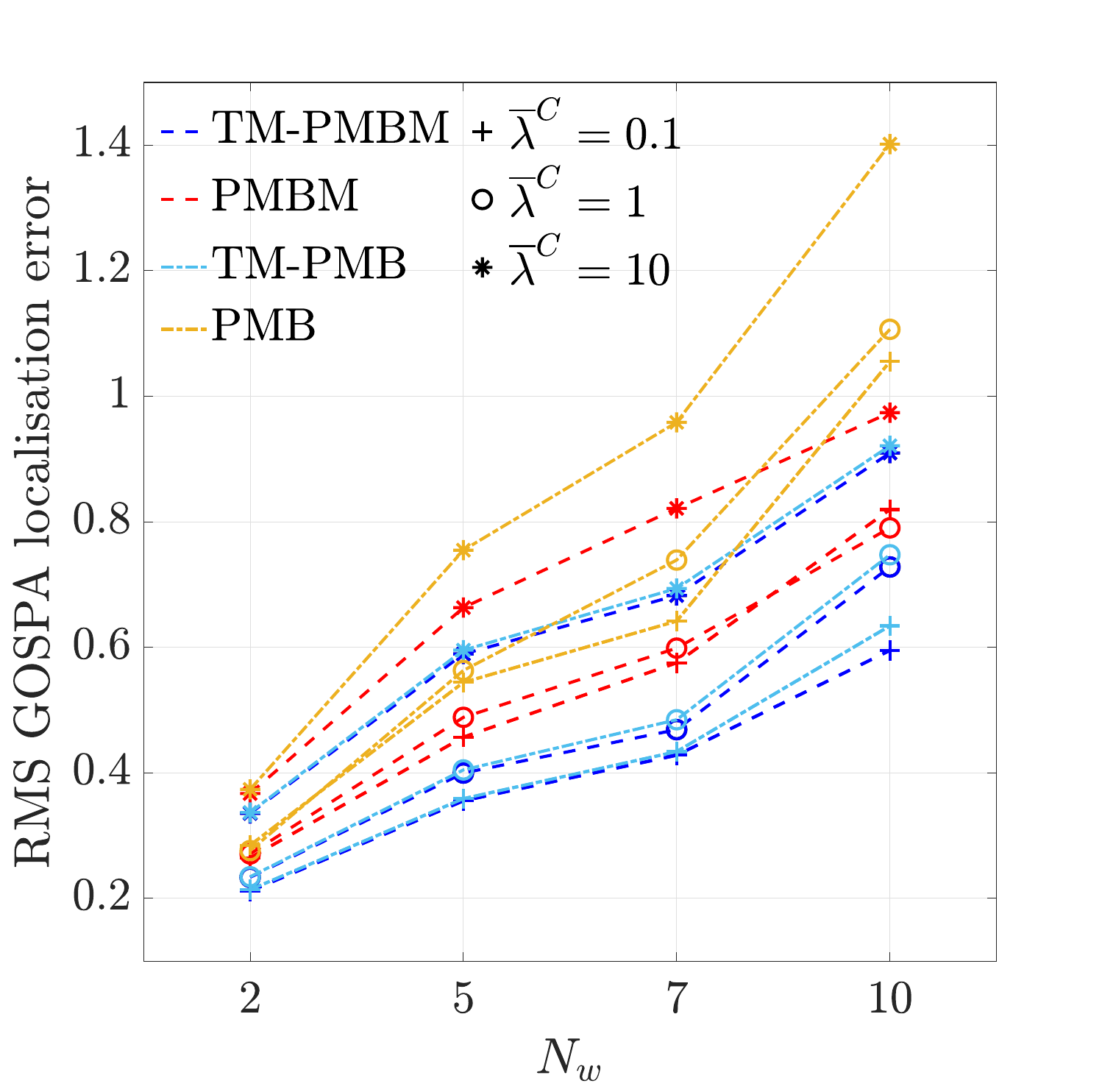}}\\
\subfloat[\label{fig:tracklets:Williams_pd07_false}]%
{\centering{}\includegraphics[scale=0.18]{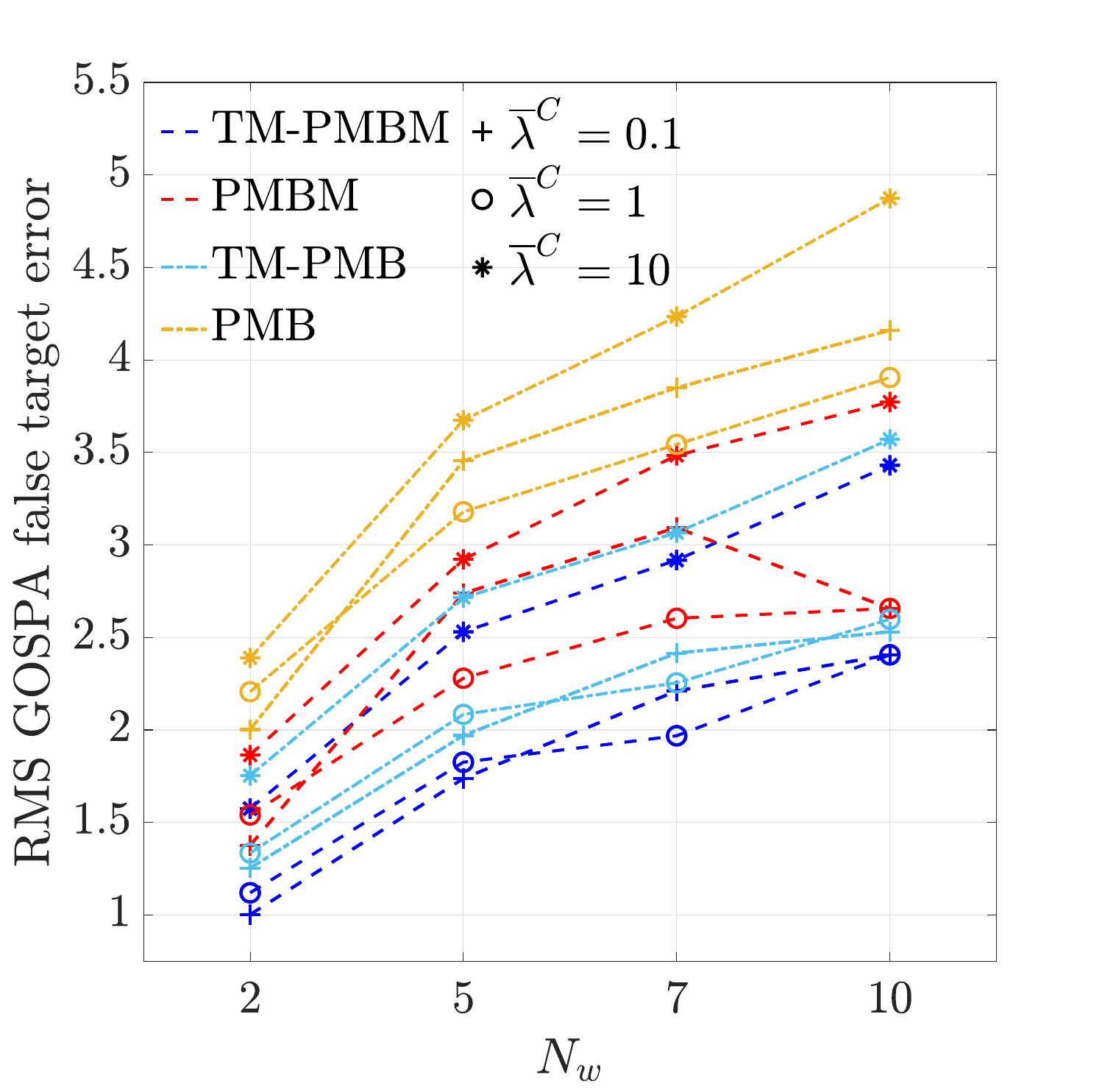}}%
\subfloat[\label{fig:tracklets:Williams_pd07_mis}]%
{\centering{}\includegraphics[scale=0.18]{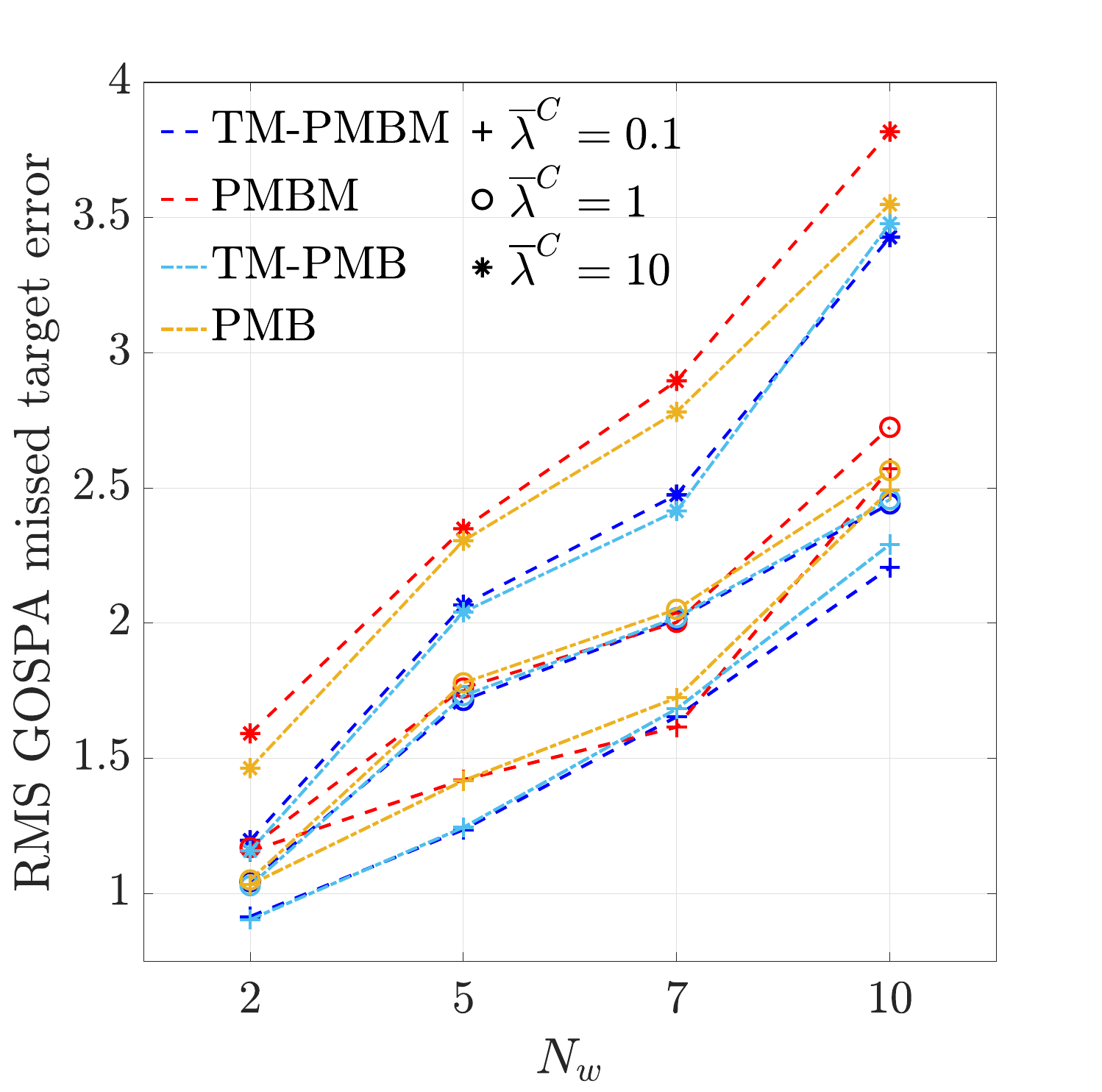}}
\par\end{centering}
\centering{}\caption{\label{fig:Results_Williams_pd07}GOSPA metric results for Scenario
1 with $\widetilde{p}^{D}=0.7$ and clutter rates $\overline{\lambda}^{C}=\{0.1,1,10\}$,
averaged over 100 MC runs. Each filter is represented by a distinct
colour, and simulations sharing the same clutter rate are identified
using the same marker. Results for the TM-PMBM and PMBM filters are
connected by dashed lines, while those for the TM-PMB and PMB filters
are connected by dash-dotted lines.}
\end{figure}

We tested this scenario with two different probabilities of receiving
a full trajectory measurement $\widetilde{p}^{D}$, specifically $\widetilde{p}^{D}\in\{0.7,0.9\}$.
The results are depicted, respectively, in Figs. \ref{fig:tracklets:Results_Williams_pd09}
and \ref{fig:Results_Williams_pd07}. For each value of $\widetilde{p}^{D}$,
we conducted three simulations with different clutter rates $\overline{\lambda}^{C}\in\{0.1,1,10\}$
for each of the following time window lengths: $N_{w}\in\left\{ 2,5,7,10\right\} $.

Both the TM-PMBM and TM-PMB filters outperform the standard PMBM and
PMB filters, respectively. Fig. \ref{fig:Comparison-single_run} presents
a comparison, based on a single MC run, of the tracking performance
of TM-PMBM and standard PMBM filters. Figs. \ref{fig:tracklets:Williams_pd09_loc}
and \ref{fig:tracklets:Willaims_pd07_loc} demonstrate that the localisation
errors tends to converge to a common value as the length of the time
window decreases. Conversely, extending the duration of the time window
results in higher localisation errors and a greater disparity in localisation
accuracy between the TM-PMBM/TM-PMB filters and the standard ones.

The missed target error is quite similar between the TM-PMBM/TM-PMB
filters and the conventional ones. However, the TM-PMBM/TM-PMB filters
significantly outperform standard filters in the high clutter scenario,
as shown in Fig. \ref{fig:tracklets:Williams_pd09_mis} for $\overline{\lambda}^{C}=10$.
Fig. \ref{fig:tracklets:Williams_pd07_mis} indicates that the difference
in missed target error generally increases with a lower $\widetilde{p}^{D}$,
especially for long time windows.

On the contrary, the TM-PMBM filter and the standard one have a similar
false target error in the high clutter scenarios, as shown in Figs.
\ref{fig:tracklets:Williams_pd09_false} for $\overline{\lambda}^{C}=10$.
Note that for PMB filters, the difference remains significant and
favours the TM-PMB filter. TM-PMBM and TM-PMB filters perform well
even in a scenario with a lower $\widetilde{p}^{D}$, as shown in
Fig. \ref{fig:tracklets:Williams_pd07_false}. However, the best performance
in the high clutter scenario is achieved by the standard PMBM.

Tab. \ref{tab:comp_times_scenario1} reports the computational times
of the filters for each simulation, with the TM-PMBM/TM-PMB filters
exhibiting lower computational times in scenarios with longer time
windows. Although the standard PMBM/PMB filters typically operate
with a larger number of single-target and global hypotheses, as showed
in the hypothesis-count analysis presented in Appendix \ref{subsec:Hypothesis-count-analysis},
the TM-PMBM/TM-PMB filters do not always outperform the standard PMBM/PMB
filters in terms of computational time, owing to the additional cost
of marginalisation and the evaluation of Gaussian distributions in
higher-dimensional spaces.
\begin{table*}
\centering{}\caption{Computational times of the TM-PMBM/TM-PMB filters and the PMBM/PMB
filters for simulations based on Scenario 1. For each scenario, the
fastest filter between TM-PMBM/PMBM and TM-PMB/PMB is underlined.
\label{tab:comp_times_scenario1}}
\begin{tabular}{c|c|cccc|cccc|cccc|cccc}
\hline 
\noalign{\vskip\doublerulesep}
\multicolumn{1}{c}{\begin{turn}{90}
\end{turn}} &
 &
\multicolumn{4}{c|}{TM-PMBM with $N_{w}=$} &
\multicolumn{4}{c|}{PMBM with $N_{w}=$} &
\multicolumn{4}{c|}{TM-PMB with $N_{w}=$} &
\multicolumn{4}{c}{PMB with $N_{w}=$}\tabularnewline
\noalign{\vskip\doublerulesep}
\multicolumn{1}{c}{\multirow{1}{*}{$\widetilde{p}^{D}$}} &
$\overline{\lambda}^{C}$ &
$2$ &
$5$ &
$7$ &
$10$ &
$2$ &
$5$ &
$7$ &
$10$ &
$2$ &
$5$ &
$7$ &
$10$ &
$2$ &
$5$ &
$7$ &
$10$\tabularnewline
\hline 
\hline 
\noalign{\vskip\doublerulesep}
\multirow{3}{*}{0.9} &
0.1 &
4.44 &
\uline{0.76} &
\uline{0.39} &
\uline{0.32} &
\uline{3.38} &
1.09 &
0.79 &
0.49 &
0.39 &
0.19 &
0.13 &
\uline{0.10} &
\uline{0.35} &
\uline{0.14} &
\uline{0.11} &
\uline{0.10}\tabularnewline
\noalign{\vskip\doublerulesep}
 & 1 &
6.15 &
\uline{1.16} &
\uline{0.60} &
\uline{0.57} &
\uline{4.48} &
1.79 &
1.18 &
0.66 &
\uline{0.45} &
0.25 &
0.17 &
0.13 &
0.51 &
\uline{0.17} &
\uline{0.16} &
\uline{0.12}\tabularnewline
\noalign{\vskip\doublerulesep}
 & 10 &
7.14 &
\uline{1.51} &
\uline{1.41} &
\uline{0.99} &
\uline{5.65} &
2.67 &
1.97 &
1.35 &
1.35 &
0.60 &
\uline{0.40} &
\uline{0.35} &
\uline{0.97} &
\uline{0.44} &
0.43 &
0.40\tabularnewline
\hline 
\noalign{\vskip\doublerulesep}
\multirow{3}{*}{0.7} &
0.1 &
3.84 &
\uline{0.91} &
\uline{0.65} &
\uline{0.40} &
\uline{3.24} &
1.14 &
0.78 &
0.55 &
0.46 &
0.20 &
\uline{0.12} &
\uline{0.10} &
\uline{0.33} &
\uline{0.13} &
\uline{0.12} &
\uline{0.10}\tabularnewline
\noalign{\vskip\doublerulesep}
 & 1 &
6.10 &
\uline{1.27} &
\uline{1.01} &
\uline{0.65} &
\uline{4.54} &
2.08 &
1.27 &
0.73 &
0.54 &
0.26 &
\uline{0.16} &
\uline{0.14} &
\uline{0.42} &
\uline{0.18} &
0.26 &
\uline{0.14}\tabularnewline
\noalign{\vskip\doublerulesep}
 & 10 &
6.67 &
\uline{2.25} &
\uline{1.29} &
\uline{1.20} &
\uline{6.31} &
3.46 &
2.50 &
1.96 &
1.22 &
0.67 &
\uline{0.47} &
\uline{0.45} &
\uline{0.99} &
\uline{0.62} &
0.75 &
0.49\tabularnewline
\hline 
\end{tabular}
\end{table*}

\subsection{Scenario 2\label{subsec:tracklets:Scenario-2}}

Scenario 2 involves targets that appear and disappear at various time
intervals within the area $A=[0,600]\times[0,400]$. The target state
at the appearing time is Gaussian with mean $[300,0,200,0]^{T}$ and
covariance $\diag([30,1,30,1]^{2})$. The birth process follows a
Poisson distribution with an average of $0.16$ targets per time step,
and the average lifespan of each target is $1000$ seconds. The scenario
has an average number of alive target at each time step $N_{a}=20.32$,
with a maximum of $45$ targets present at the final time step. The
PPP intensity at each is Gaussian with mean $[300,0,200,0]^{T}$ and
covariance $\diag([(300,1,200,1]^{2})$.

\begin{figure}
\begin{centering}
\subfloat[$\widetilde{p}^{D}=0.9$\label{fig:tracklets:Real_pd09_tot}]%
{\centering{}\includegraphics[scale=0.18]{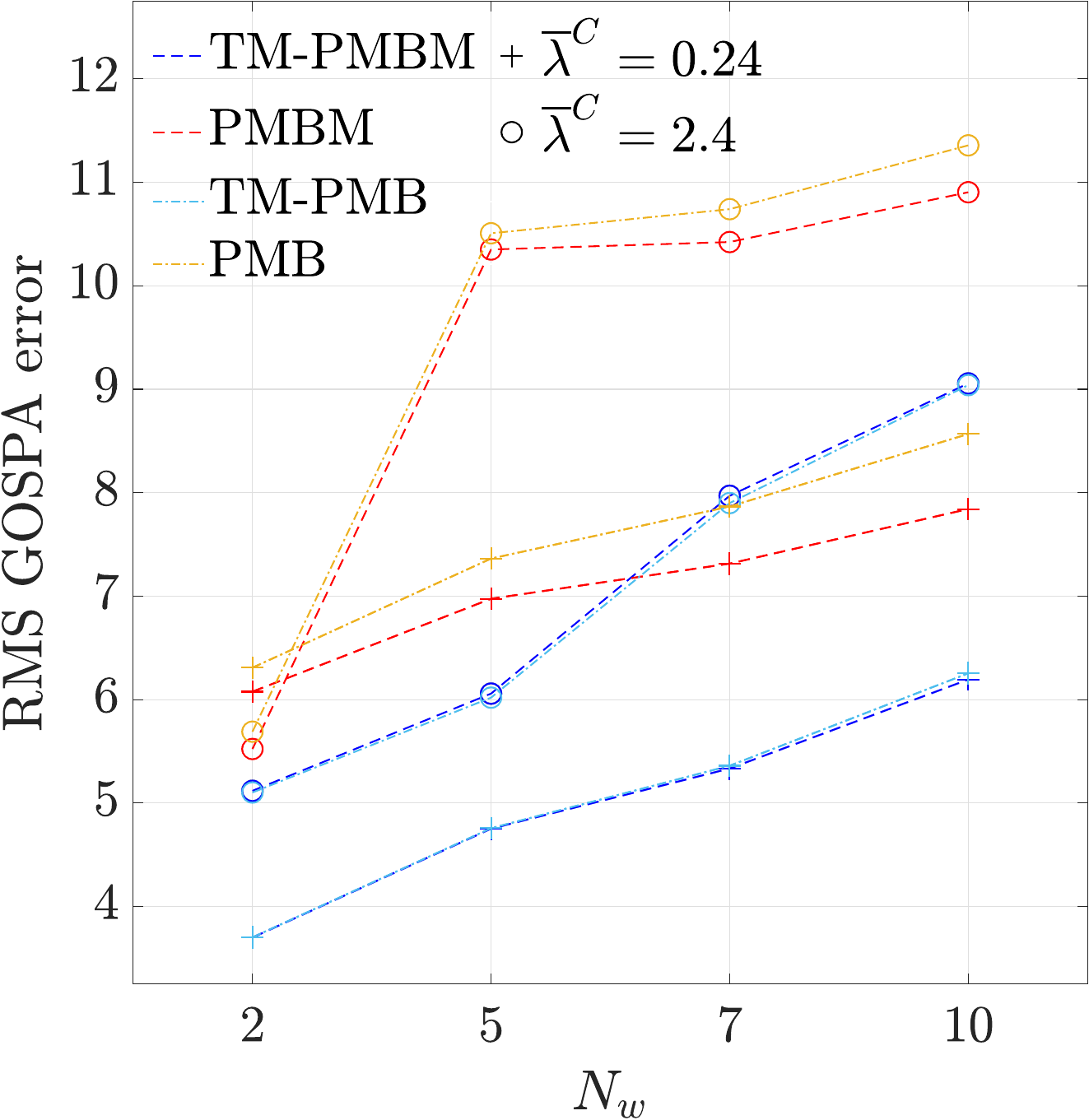}}\quad{}%
\subfloat[$\widetilde{p}^{D}=0.7$\label{fig:tracklets:Real_pd07_tot}]%
{\centering{}\includegraphics[scale=0.18]{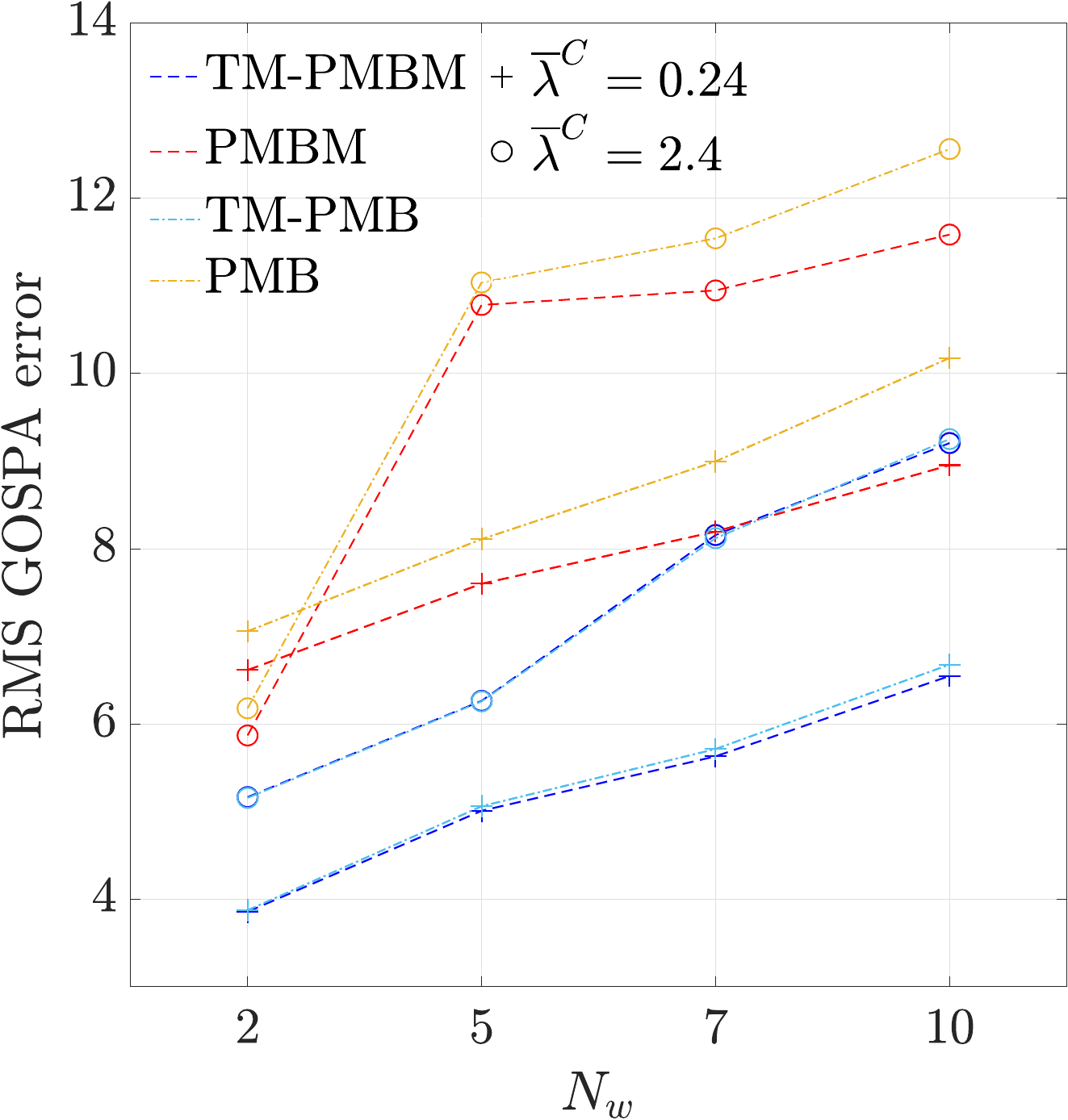}}
\par\end{centering}
\centering{}\caption{\label{fig:tracklets:Results_Real}GOSPA metric results for Scenario
2 with $\widetilde{p}^{D}=\{0.9,0.7\}$ and clutter rates $\overline{\lambda}^{C}=\{0.24,2.4\}$,
averaged over 100 Monte Carlo runs. Each filter is represented by
a distinct colour, and simulations sharing the same clutter rate are
identified using the same marker.}
\end{figure}

Fig. \ref{fig:tracklets:Results_Real} presents the results in terms
of RMS GOSPA error with $\widetilde{p}^{D}\in\{0.9,0.7\}$ and clutter
rates $\overline{\lambda}^{C}\in\{0.24,2.4\}$. In both configurations,
the performance of the TM-PMB filter is much more similar to that
of the TM-PMBM one compared to the conventional filters. Overall,
TM-PMBM and TM-PMB outperform their corresponding versions based on
target states.

\section{Conclusions\label{sec:tracklets:Conclusions}}

In this paper, we have proposed a PMBM filter that processes sets
of sensor measurements, in which each measurement represents a trajectory
in a two-time step window. We have derived the filtering recursion
by first performing the prediction step on the PMBM posterior density
on the target states, followed by the update step with a set of trajectory
measurements, which requires the use of the PMBM on the set of trajectories
over the last two time steps. We have then marginalised the PMBM density
to obtain a posterior on the target state estimation accuracy.

We have also introduced the TM-PMB approximation and a Gaussian implementation
of the TM-PMBM filter suitable for linear/Gaussian measurement models.
Finally, we compared the performance of the TM-PMBM and TM-PMB filters
with their counterparts based on target states in two scenarios. The
TM-PMBM and TM-PMB filters demonstrated superior performance in both
scenarios.

\section*{Acknowledgements}

The authors thank Sintela Ltd. and the EPSRC Centre for Doctoral Training
in Distributed Algorithms for their support. Simon Maskell's work
was supported by the Royal Academy of Engineering. The authors also
thank Paul Horridge for his helpful discussions on a closely related
issue.

\bibliographystyle{IEEEtran}
\bibliography{16C__Users_marfon_OneDrive_-_The_University_of_Liverpool_Marco_Fontana}

\cleardoublepage{}

{\LARGE Supplementary material: Poisson multi-Bernoulli mixture filter
for trajectory measurements}{\LARGE\par}

\appendices{}

\section{\label{sec:Proof-of-Lemma5}}

In this appendix, we prove the expression of the updated Bernoulli
components in Lemma \ref{lem:Gaus_TMPMBM_upd}. For clarity, we simplify
the notation in this section by omitting the dependence on $k+1$
of the single-trajectory measurements, single-trajectory state and
related spaces.

\subsection{Update of Bernoulli component\label{subsec:supp:Update-of-Bernoulli}}

According to Lemma \ref{lem:Gaus_TMPMBM_pred}, the predicted Bernoulli
component $f_{k+1|k}^{i,\widetilde{a}^{i}}(\cdot)$, $i\in\{1,\dots,n_{k+1|k}\}$,
$\widetilde{a}^{i}\in\{1,\dots,h_{k+1|k}^{i}\}$ has single-target
density of the form (\ref{eq:Gaus_single_trj_density}). Based on
Lemma \ref{lem:TM-PMBM update}, the detection hypothesis for $f_{k+1|k}^{i,\widetilde{a}^{i}}(\cdot)$
and trajectory measurement $Z^{j}=(t,z_{1:\iota})$ is denoted as
$f_{k+1|k}^{i,\widetilde{a}^{i}}(\cdot)$ with index $a^{i}=\widetilde{a}^{i}+h_{k+1|k}^{i}j$
and has parameters that depend on the inner product
\begin{align}
\left\langle \right. & \left.p_{k+1|k}^{i,\widetilde{a}^{i}},l\left(Z_{k+1}^{j}|\cdot\right)p^{D}\right\rangle \nonumber \\
 & =p^{D}\int_{T^{1}\uplus T^{3}}l\left(Z^{j}|X\right)p_{k+1|k}^{i,\widetilde{a}^{i}}\left(X\right)dX\label{eq:}\\
 & =p^{D}\left[\beta_{k+1|k}^{i,a^{i}}(1)\right.\nonumber \\
 & \quad\times\int_{T^{1}}l\left(Z^{j}|X\right)\mathcal{N}(X;k,\overline{x}_{k+1|k}^{i,\widetilde{a}^{i}}(1),P_{k+1|k}^{i,\widetilde{a}^{i}}(1))dX\nonumber \\
 & \quad+\beta_{k+1|k}^{i,a^{i}}(2)\nonumber \\
 & \left.\quad\times\int_{T^{3}}l\left(Z^{j}|X\right)\mathcal{N}(X;k,\overline{x}_{k+1|k}^{i,\widetilde{a}^{i}}(2),P_{k+1|k}^{i,\widetilde{a}^{i}}(2))dX\right].\label{eq:generic_inner_product}
\end{align}
Note that the integral does not consider $T^{2}$ as these single-trajectory
densities do not consider new born targets at time step $k+1$. We
derive (\ref{eq:generic_inner_product}) for each possible trajectory
measurement defined in (\ref{eq:def_tracklet}).

\subsubsection{Measurements detected only at time step $k$\label{subsec:supp:Measurements-detected-only}}

If $Z^{j}=(k,z_{1})\in M^{1}$, the measurement model takes the form
of (\ref{eq:l_case2}) and (\ref{eq:l_case5}) for $X\in T^{3}$ and
$X\in T^{1}$, respectively. Therefore, (\ref{eq:generic_inner_product})
yields two non-zero terms 
\begin{align}
\left\langle \right. & \left.p_{k+1|k}^{i,\widetilde{a}^{i}},l\left(Z_{k+1}^{j}|\cdot\right)p^{D}\right\rangle \nonumber \\
 & =p^{D}\left[\beta_{k+1|k}^{i,a^{i}}(1)\right.\nonumber \\
 & \quad\times\int_{\mathbb{R}^{n_{x}}}h_{1,1}(z_{1}|x_{1})\mathcal{N}(x_{1},\overline{x}_{k+1|k}^{i,\widetilde{a}^{i}}(1),P_{k+1|k}^{i,\widetilde{a}^{i}}(1))dx_{1}\nonumber \\
 & \quad+\beta_{k+1|k}^{i,a^{i}}(2)\gamma\nonumber \\
 & \left.\quad\times\int_{\mathbb{R}^{2n_{x}}}h_{1,3}(z_{1}|x_{1:2})\mathcal{N}(x_{1:2},\overline{x}_{k+1|k}^{i,\widetilde{a}^{i}}(2),P_{k+1|k}^{i,\widetilde{a}^{i}}(2))dx_{1:2}\right]\label{eq:-1}\\
 & =p^{D}\left[\beta_{k+1|k}^{i,a^{i}}(1)\right.\nonumber \\
 & \quad\times\int_{\mathbb{R}^{n_{x}}}\mathcal{N}\left(z_{1};H_{1,1}x_{1},R_{1}\right)\mathcal{N}(x_{1},\overline{x}_{k+1|k}^{i,\widetilde{a}^{i}}(1),P_{k+1|k}^{i,\widetilde{a}^{i}}(1))dx_{1}\nonumber \\
 & \quad+\beta_{k+1|k}^{i,a^{i}}(2)\gamma\int_{\mathbb{R}^{2n_{x}}}\mathcal{N}\left(z_{1};H_{1,3}x_{1:2},R_{1}\right)\nonumber \\
 & \left.\quad\times\mathcal{N}(x_{1:2},\overline{x}_{k+1|k}^{i,\widetilde{a}^{i}}(2),P_{k+1|k}^{i,\widetilde{a}^{i}}(2))dx_{1}dx_{2}\right]\label{eq:-2}\\
 & =p^{D}\left[\beta_{k+1|k+1}^{i,a^{i}}(1)\mathcal{N}\left(z_{1};\overline{z}^{i,a^{i}}(1),S_{i,a^{i}}(1)\right)\right.\nonumber \\
 & \quad\left.+\beta_{k+1|k+1}^{i,a^{i}}(2)\mathcal{N}\left(z_{1};\overline{z}^{i,a^{i}}(2),S_{i,a^{i}}(2)\right)\right]\label{eq:inner_prod_M1}
\end{align}
where
\begin{align}
\overline{z}^{i,a^{i}}(l) & =H(l)\overline{x}_{k+1|k}^{i,\widetilde{a}^{i}}(l)\label{eq:-3}\\
S_{i,a^{i}}(l) & =H(l)P_{k+1|k}^{i,\widetilde{a}^{i}}(l)H^{T}(l)+R\label{eq:-4}
\end{align}
with $l\in\{1,2\}$, $H(1)=H_{1,1}=H$ and $H(2)=H_{1,3}=[1,0]\otimes H$.
Therefore, $\overline{z}^{i,a^{i}}(1)=\overline{z}^{i,a^{i}}(2)$,
$S_{i,a^{i}}(1)=S_{i,a^{i}}(2)$ and (\ref{eq:inner_prod_M1}) yields
(\ref{eq:weight_died_upd}). Based on (\ref{eq:single_state_density_gen_upd_det-1}),
the Kalman update gives the parameters for the updated single-trajectory
density of the form (\ref{eq:Gaus_single_trj_density}).

\subsubsection{Measurement detected at both time steps $k$ and $k+1$\label{subsec:supp:Measurement-detected-at}}

If $Z^{j}=(k,z_{1:2})\in M^{3}$, the measurement model takes the
form of (\ref{eq:l_case1}) for $X\in T^{3}$, while it is zero for
$X\in T^{1}$. Therefore, (\ref{eq:generic_inner_product}) yields
a non-zero term corresponding to $l=2$ in (\ref{eq:Gaus_single_trj_density})
\begin{align}
\left\langle \right. & \left.p_{k+1|k}^{i,\widetilde{a}^{i}},l\left(Z_{k+1}^{j}|\cdot\right)p^{D}\right\rangle \nonumber \\
 & =p^{D}\beta_{k+1|k}^{i,a^{i}}(2)\widetilde{p}^{D}\nonumber \\
 & \quad\times\int_{\mathbb{R}^{2n_{x}}}h_{3,3}(z_{1:2}|x_{1:2})\mathcal{N}(x_{1:2},\overline{x}_{k+1|k}^{i,\widetilde{a}^{i}}(2),P_{k+1|k}^{i,\widetilde{a}^{i}}(2))dx_{1:2}\label{eq:-5}\\
 & =p^{D}\beta_{k+1|k}^{i,a^{i}}(2)\widetilde{p}^{D}\int_{\mathbb{R}^{2n_{x}}}\mathcal{N}\left(z_{1:2};H_{3,3}x_{1:2},R_{2}\right)\nonumber \\
 & \quad\times\mathcal{N}(x_{1:2},\overline{x}_{k+1|k}^{i,\widetilde{a}^{i}}(2),P_{k+1|k}^{i,\widetilde{a}^{i}}(2))dx_{1:2}\label{eq:-6}\\
 & =p^{D}\beta_{k+1|k+1}^{i,a^{i}}(2)\mathcal{N}\left(z_{1:2};\overline{z}^{i,a^{i}},S_{i,a^{i}}\right),\label{eq:inner_prod_M3}
\end{align}
where 
\begin{align}
\overline{z}^{i,a^{i}} & =H_{3,3}\overline{x}_{k+1|k}^{i,\widetilde{a}^{i}}(2)\label{eq:-7}\\
S_{i,a^{i}} & =H_{3,3}P_{k+1|k}^{i,\widetilde{a}^{i}}(2)H_{3,3}^{T}+R_{2},\label{eq:-8}
\end{align}
with $H_{3,3}=I_{2}\otimes H$ and $R_{2}=I_{2}\otimes R$. The expression
for the weight (\ref{eq:weigth_alive_upd}) follows directly from
(\ref{eq:inner_prod_M3}). Based on (\ref{eq:single_state_density_gen_upd_det-1}),
the Kalman update gives the parameters for the updated single-trajectory
density of the form (\ref{eq:Gaus_single_trj_density}).

\subsubsection{Measurement detected only at time step $k+1$\label{subsec:supp:Measurement-detected-only}}

If $Z^{j}=(k+1,z_{1})\in M^{2}$, the measurement model takes the
form of (\ref{eq:l_case3}) for $X\in T^{3}$, while it is zero for
$X\in T^{1}$. Therefore, (\ref{eq:generic_inner_product}) yields
a non-zero term corresponding to $l=2$ in (\ref{eq:Gaus_single_trj_density})
\begin{align}
\left\langle \right. & \left.p_{k+1|k}^{i,\widetilde{a}^{i}},l\left(Z_{k+1}^{j}|\cdot\right)p^{D}\right\rangle \nonumber \\
 & =p^{D}\beta_{k+1|k}^{i,a^{i}}(2)\gamma\nonumber \\
 & \quad\times\int_{\mathbb{R}^{2n_{x}}}h_{2,3}(z_{1}|x_{1:2})\mathcal{N}(x_{1:2},\overline{x}_{k+1|k}^{i,\widetilde{a}^{i}}(2),P_{k+1|k}^{i,\widetilde{a}^{i}}(2))dx_{1:2}\label{eq:-11}\\
 & =p^{D}\beta_{k+1|k}^{i,a^{i}}(2)\gamma\int_{\mathbb{R}^{2n_{x}}}\mathcal{N}\left(z_{1};H_{2,3}x_{1:2},R_{1}\right)\nonumber \\
 & \quad\times\mathcal{N}(x_{1:2},\overline{x}_{k+1|k}^{i,\widetilde{a}^{i}}(2),P_{k+1|k}^{i,\widetilde{a}^{i}}(2))dx_{1:2}\label{eq:-12}\\
 & =p^{D}\beta_{k+1|k+1}^{i,a^{i}}(2)\mathcal{N}\left(z_{1};\overline{z}^{i,a^{i}},S_{i,a^{i}}\right),\label{eq:inner_prod_M2}
\end{align}
where 
\begin{align}
\overline{z}^{i,a^{i}} & =H_{2,3}\overline{x}_{k+1|k}^{i,\widetilde{a}^{i}}(2)\label{eq:-9}\\
S_{i,a^{i}} & =H_{2,3}P_{k+1|k}^{i,\widetilde{a}^{i}}(2)H_{2,3}^{T}+R_{2},\label{eq:-10}
\end{align}
with $H_{2,3}=[0,1]\otimes H$ and $R_{2}=I_{2}\otimes R$. The expression
for the weight (\ref{eq:weigth_alive_upd}) follows directly from
(\ref{eq:inner_prod_M3}). Based on (\ref{eq:single_state_density_gen_upd_det-1}),
the Kalman update gives the parameters for the updated single-trajectory
density of the form (\ref{eq:Gaus_single_trj_density}).

\subsection{New Bernoulli component}

According to Lemma \ref{lem:Gaus_TMPMBM_pred}, the predicted intensity
is a Gaussian mixture intensity comprising two different kinds of
components, denoted as $b$ and $p$ in (\ref{eq:PPP_GM}). Based
on Lemma \ref{lem:TM-PMBM update}, the detection hypothesis of a
new Bernoulli component $f_{k+1|k+1}^{i,a^{i}}(\cdot)$, $i\in\{n_{k+1|k}+j\}$,
$j\in\{1,\dots,m_{k+1}\}$, initiated by the trajectory measurement
$Z^{j}=(t,z_{1:\nu})\in M^{\mu}$is computed by evaluating the inner
product $\left\langle \lambda_{k+1|k},l\left(Z^{j}|\cdot\right)p^{D}\right\rangle $.

For the $q_{b}$-th component of the mixture $b$ in (\ref{eq:PPP_GM}),
the measurement model is non-zero only if $Z^{j}=(k+1,z_{1})\in M^{2}$
and takes the form of (\ref{eq:l_case9}). Therefore, 
\begin{align}
v^{b,q_{b}} & =\left\langle \lambda_{k+1|k},l\left(Z^{j}|\cdot\right)p^{D}\right\rangle \nonumber \\
 & =w_{k+1|k}^{b,q_{b}}p^{D}\int_{\mathbb{R}^{n_{x}}}h_{2,2}(z_{1}|x_{1})\mathcal{N}(x_{1},\overline{x}_{k+1|k}^{b,q_{b}},P_{k+1|k}^{b,q_{b}})dx_{1}\label{eq:-13}\\
 & =w_{k+1|k}^{b,q_{b}}p^{D}\nonumber \\
 & \quad\times\int_{\mathbb{R}^{n_{x}}}\mathcal{N}(z_{1},Hx_{1},R_{1})\mathcal{N}(x_{1},\overline{x}_{k+1|k}^{b,q_{b}},P_{k+1|k}^{b,q_{b}})dx_{1}\label{eq:-14}\\
 & =w_{k+1|k}^{b,q_{b}}p^{D}\mathcal{N}(z_{1},H\overline{x}_{k+1|k}^{b,q_{b}},S_{b,q_{b}}),\label{eq:v_b_qb}
\end{align}
where $S_{b,q_{b}}=HP_{k+1|k}^{b,q_{b}}H^{T}+R$. For the $q_{p}$-th
component of the mixture $p$ in (\ref{eq:PPP_GM}), the measurement
model is in the forms of (\ref{eq:l_case1})-(\ref{eq:l_case3}),
and the correspondent inner product $v^{p,q_{p}}$ can be computed
similarly to (\ref{eq:inner_prod_M1})-(\ref{eq:inner_prod_M2}).

Once the inner products are computed, (\ref{eq:W_newB_Gauss}) and
(\ref{eq:R_newB_Gauss}) follow directly from (\ref{eq:tracklets:w_gen_upd_new})
and (\ref{eq:r_gen_upd_new-1}). Finally, from (\ref{eq:single_state_density_gen_upd_new-1}),
the single-trajectory density of the new Bernoulli component is a
Gaussian mixture, where each component results from a Kalman update
based on (\ref{eq:v_b_qb}) or (\ref{eq:tracklets:v_p}), similar
to the Bernoulli competent updates in Sec. \ref{subsec:supp:Update-of-Bernoulli}.
In this paper, we propose to use a Gaussian approximation of the resulting
mixture (\ref{eq:Gaussian_state_update_newB}).

\section{Hypothesis-count analysis\label{subsec:Hypothesis-count-analysis}}

\begin{figure}[h]
\begin{centering}
\subfloat[Mean number of single-target hypotheses per time step, averaged over
30 MC runs, for the TM-PMBM and PMBM filters.\label{fig:tracklets:PMBM_loc_hypo_count}]%
{\centering{}\includegraphics[scale=0.18]{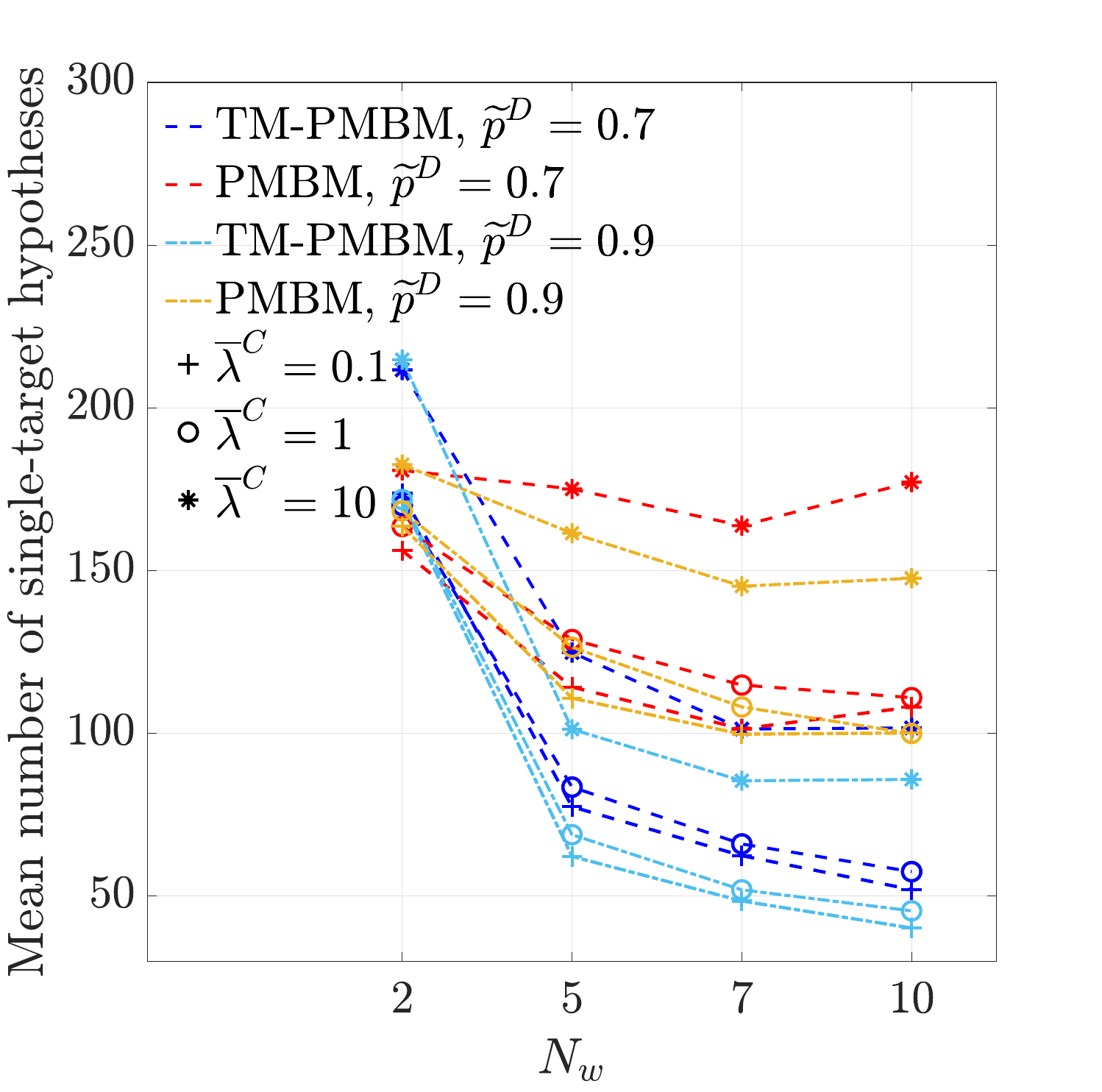}}%
\subfloat[Mean number of global hypotheses per time step, averaged over 30 MC
runs, for the TM-PMBM and PMBM filters.\label{fig:tracklets:PMBM_glob_hypo_count}]%
{\centering{}\includegraphics[scale=0.18]{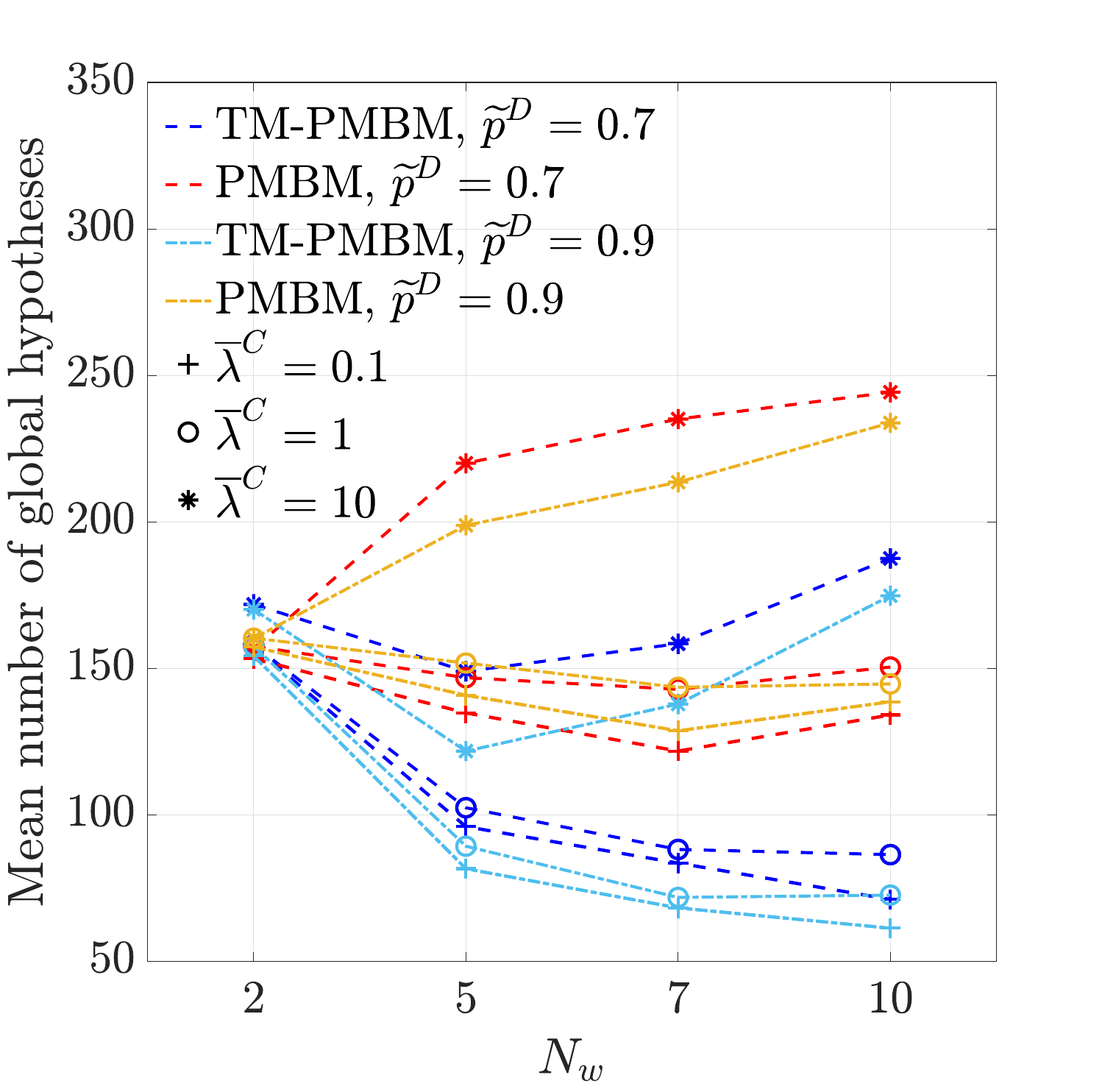}}
\par\end{centering}
\centering{}%
\subfloat[Mean number of single-target hypotheses per time step, averaged over
30 MC runs, for the TM-PMB and PMB filters.\label{fig:PMB_tracklets:loc_hypo_count}]%
{\centering{}\includegraphics[scale=0.18]{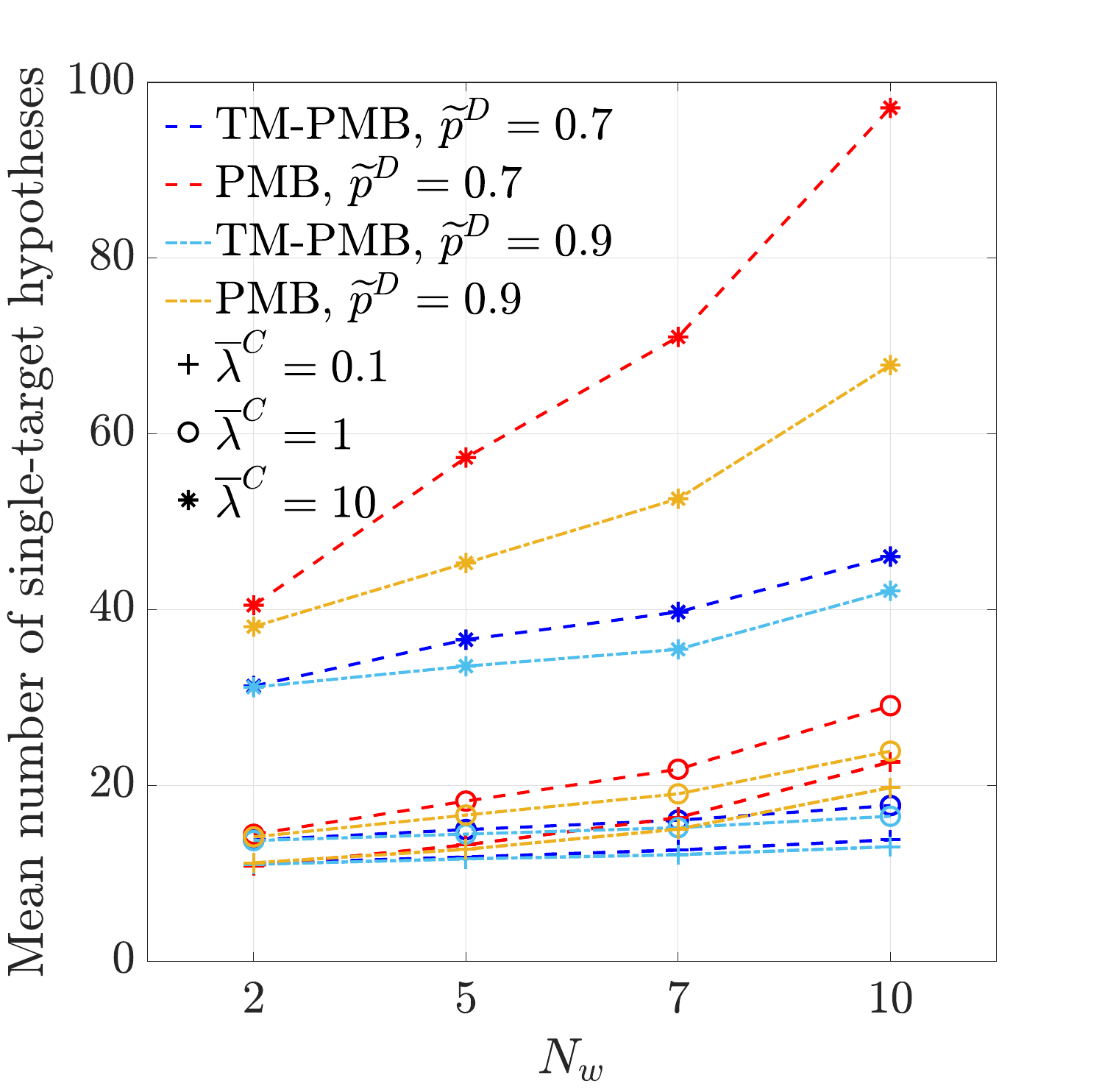}}%
\subfloat[Mean number of global hypotheses per time step, averaged over 30 MC
runs, for the TM-PMB and PMB filters.\label{fig:tracklets:PMB_glob_hypo_count}]%
{\centering{}\includegraphics[scale=0.18]{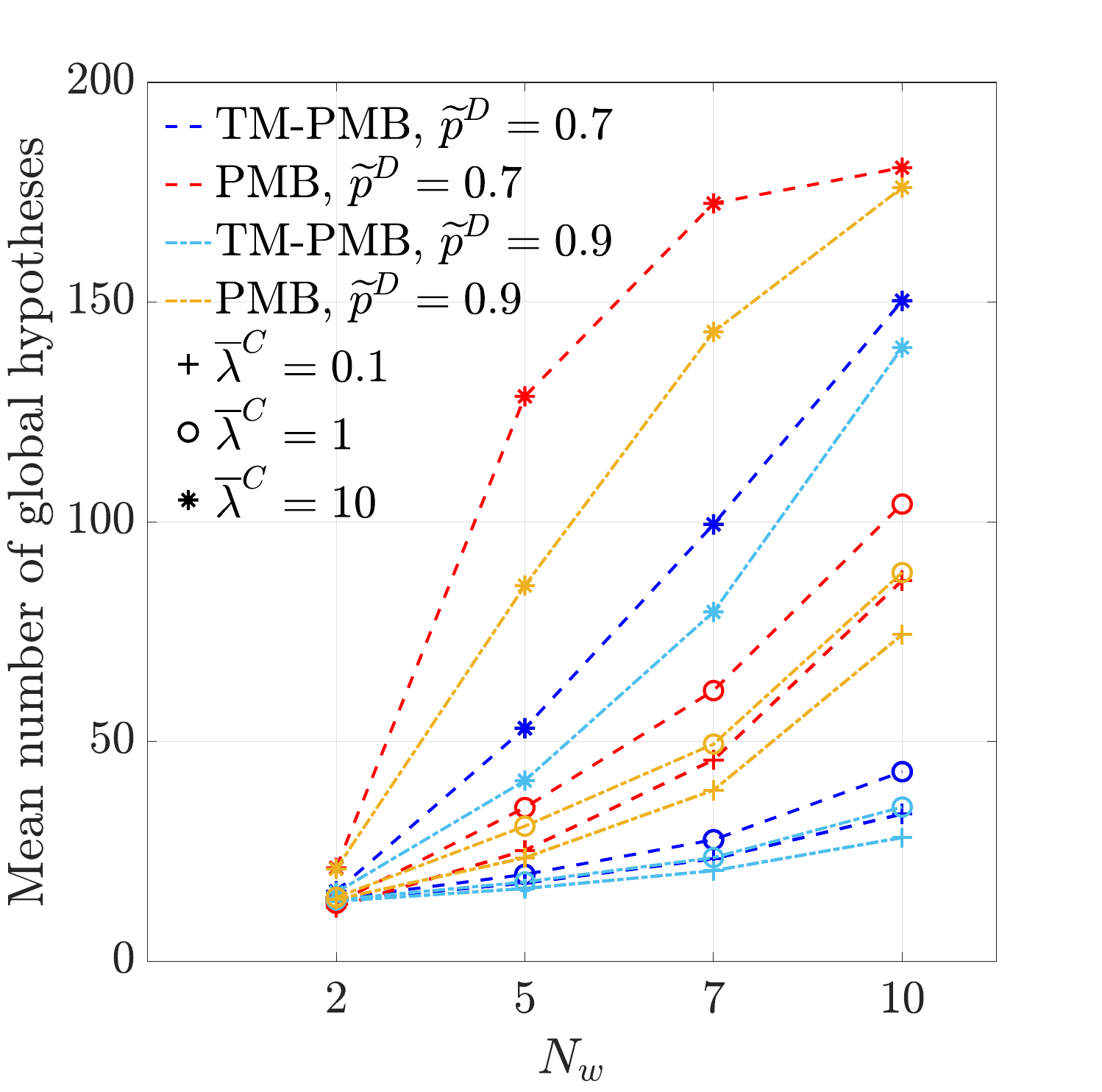}}\caption{Hypothesis-count analysis for simulations in Scenario 1.\label{fig:tracklets:hypo_count}}
\end{figure}

In this appendix, we present a hypothesis-count analysis of the TM-PMBM/TM-PMB
and PMBM/PMB filters to justify the computational times in Tab. \ref{tab:comp_times_scenario1}.
Fig. \ref{fig:tracklets:hypo_count} shows the average number of single-target
hypotheses and global hypotheses in the filters at each time step
(or two-time step window), for the simulations based on Scenario 1
in Sec. \ref{subsec:tracklets:Scenario-1}. Apart from the case with $N_{w}=2$, Fig. \ref{fig:tracklets:hypo_count}
shows that the TM-PMBM/TM-PMB filters operate with fewer single-target
and global hypotheses than the standard PMBM/PMB filters.

\end{document}